\documentclass[a4paper,10.5pt]{article}
\usepackage{amsfonts}
\usepackage{amsmath}
\usepackage{epsfig}
\usepackage{latexsym}
\usepackage{amssymb}
\usepackage{color}
\usepackage{amscd}
\usepackage{multirow}
\usepackage{graphicx}
\usepackage{lscape}
\usepackage{bm}
\usepackage{natbib}
\usepackage[ruled,vlined]{algorithm2e}

\newcommand{\Remm}[1]{}

\newtheorem{model ass}[theo]{Model Assumptions}

\numberwithin{equation}{section}

\begin{document}

\noindent{\large {\textbf{Bayesian Cointegrated Vector Autoregression models incorporating $\alpha$-stable noise for inter-day price movements via Approximate Bayesian Computation.}}}

\vspace{0.2cm} \noindent\textbf{Gareth W.~Peters,}~({\footnotesize{\textit{Corresponding author}}})\newline
{\footnotesize {UNSW Mathematics and Statistics Department, Sydney, 2052, Australia;\newline
email: garethpeters@unsw.edu.au}}

\vspace{0.05cm} \noindent\textbf{Balakrishnan B. Kannan,~ Ben Lasscock,~ Chris Mellen} \newline
{\footnotesize Boronia Capital Pty. Ltd., 12 Holtermann St., Crows Nest, NSW 2065, Australia.}

\vspace{0.05cm} \noindent\textbf{Simon ~Godsill} \newline
{\footnotesize Department of Electrical Engineering, University of Cambridge, UK.}

\section{Abstract}
We consider a statistical model for pairs of traded assets, based on a Cointegrated Vector Auto Regression (CVAR) Model. We extend standard CVAR models to incorporate estimation of model parameters in the presence of price series level shifts which are not accurately modeled in the standard Gaussian error correction model (ECM) framework. This involves developing a novel matrix variate Bayesian CVAR mixture model comprised of Gaussian errors intra-day and $\alpha$-stable errors inter-day in the ECM framework. To achieve this we derive a novel conjugate posterior model for the Scaled Mixtures of Normals (SMiN CVAR) representation of $\alpha$-stable inter-day innovations. These results are generalized to asymmetric models for the innovation noise at inter-day boundaries allowing for skewed $\alpha$-stable models. 

Our proposed model and sampling methodology is general, incorporating the current literature on Gaussian models as a special subclass and also allowing for price series level shifts either at random estimated time points or known \textit{a priori} time points. We focus analysis on regularly observed non-Gaussian level shifts that can have significant effect on estimation performance in statistical models failing to account for such level shifts, such as at the close and open of markets. We compare the estimation accuracy of our model and estimation approach to standard frequentist and Bayesian procedures for CVAR models when non-Gaussian price series level shifts are present in the individual series, such as inter-day boundaries.  We fit a bi-variate $\alpha$-stable model to the inter-day jumps and model the effect of such jumps on estimation of matrix-variate CVAR model parameters using the likelihood based Johansen procedure and a Bayesian estimation. 
We illustrate our model and the corresponding estimation procedures we develop on both synthetic and actual data.
\\
\\
{\small{\textbf{Keywords:} Cointegrated Vector Autoregression, $\alpha$-stable, Approximate Bayesian Computation.}}\\
Journal of Econometrics
\newpage
\section{Introduction}

\noindent 

In this paper we consider accurate estimation of statistical models for pairs trading strategies. This is significant since recent empirical studies by \cite{bock2009regime} and \cite{gatev2006pairs} have shown, that in spite of the increasing volume of statistical arbitrage quantitative funds performing algorithmic trading, statistical pair trading still seems to be consistently assessed as a profitable trading strategy, providing motivation to further develop such models. 

We focus on cointegrated vector autoregression (CVAR) models which have been studied widely in the econometric literature, see \cite{engle1987co}, \cite{sugita2009monte}. For the error correction representation of a co-integrated series, see \cite{granger2001time} and the overview of \cite{koop2006bayesian}. Bayesian analysis of CVAR models has been addressed in several papers, see \cite{bauwens1994identification}, \cite{geweke1996bayesian}, \cite{kleibergen2009shape}, \cite{ackert1999stochastic} and \cite{Sugita2002}. In practice, Bayesian and non-Bayesian CVAR models are used extensively in pairs trading, see an example in \cite{peters2010model}. We demonstrate that when estimating even the basic CVAR models using data which is sampled at a frequency less than one day, on real price series pairs, the accuracy and robustness of the statistical model fit and estimation and therefore the stability of the selected portfolio weights, is strongly affected by level shifts or jumps in price series due to inter-day movements. This is evident in settings in which the cointegration rank is assumed known and so would be compounded in settings in which uncertainty in the rank is also assumed, see analysis in \cite{sugita2009monte}.

Level shifts in each price series are due to complicated economic and social market factors, we do not attempt to explain these with an economic rationale in this paper. Instead we demonstrate firstly that they occur regularly at the open and close of markets between joint trading times for pairs and secondly that statistical inference based on data that fails to appropriately account for these level shifts in a co-integration framework will result in poor model calibrations. We then develop and demonstrate a robust statistical approach to overcome this practically important estimation problem.

Typically one observes level shifts in the price series occurring as a result of the time delay between the open and close of markets for each asset in the traded pair. However, the level of the price shifts can not solely be accounted for by the evolution of the statistical model during the time period in which either market is closed. Some asset pairs may only have short periods of overlap in which each market is open and therefore the joint assets can be traded, it is particularly important to accurately model the inter-day level shifts for such pairs. We demonstrate that one can not ignore this practical issue of price series level shifts as it can result in significant sensitivity in the estimated model parameters. This in turn has consequences for trading resulting from the knowledge of the cointegration deviation series, which is affected and therefore results in carry on effects for design of trading thresholds.

We begin by studying the statistical properties of these inter-day level shifts in the differenced price series for several pairs of assets over multiple contract segments spanning several years. Each pair is chosen as they demonstrate historically statistically significant cointegration properties. We model the level shifts in each price series via the flexible class of $\alpha$-stable models, see \cite{zolotarev86}, \cite{samorodnitsky+g94}, \cite{qiou+r98b} and \cite{nolan97}. This class of models is of particular interest as they are flexible in terms of skew and kurtosis, whilst also admitting Gaussian distributions as a family member. That is, we fit the $\alpha$-stable models to the price level shift, obtained between the open and close of the time when both markets are trading. We demonstrate that in most cases the assumption of Gaussian residuals for these time periods, implied by fitting the basic CVAR model is inadequate. In particular several assets demonstrate significantly heavy tailed distributions are appropriate for capturing the inter-day price deviations resulting from these level shifts. Therefore this contradicts the typical statistical assumption, of constant homoskedastic multi-variate Gaussian innovation noise, made when fitting the basic CVAR models that are widely utilized when trading pairs or assets. As a consequence we propose a new CVAR model and Bayesian estimation framework to incorporate the potential for a $\alpha$-stable innovation noise at these particular known, deterministic time points. Thereby reducing the sensitivity of the estimated CVAR model parameters to the period in which both asset markets are not active. This can trivially be extended to include stochastic time points in a change point or switching structure. 

This differs from the work of \cite{chen2010subsampling} which develops a cointegration model for pairs of assets in which only symmetric $\alpha$-stable innovations are utilized at all trading time points, with a fixed tail index parameter $\alpha$ throughout the time series. We argue that this is an overly restrictive model simplification when used for trading purposes and in addition their approach can not be easily generalized to a Bayesian estimation framework, in which we focus our statistical estimation methodology. Their approach generalizes the Johannsen procedure \cite{johansen1990maximum} to the symmetric $\alpha$-stable innovation setting for testing the rank of the cointegrated VAR model. We will demonstrate a more flexible model removing the symmetry assumption for the stable noise, introducing a more realistic mixture noise model and providing a novel Approximate Bayesian Computation (ABC) sampling methodology for estimation and rank selection, generalizing the approach of \cite{peters2010model}. 

Estimation of the matrix variate parameters of a CVAR model under either a Johansen based likelihood-procedure or a Bayesian modeling approach will be demonstrated, on both synthetic and actual data, to be adversely affected by level shifts in the price series occurring at the close and open of markets. This will typically be reflected in large changes in the estimated CVAR model parameters, especially the constant mean level, the cointegration vectors and noise covariance matrix. In such situations, trading systems utilizing such parameter estimates will therefore also be sensitive to the changes in parameter estimates arising from the level shifts at day break boundaries. In high-frequency settings, where estimations are performed anywhere between several seconds to 20 - 30 min intervals, simply discarding the time periods during which level shifts occur can result in significant loss of trading activity. This is especially the case when trading activity is occuring around close and open times of markets. In addition, when modeling in the setting in which level shifts can occur randomly throughout the trading day, discarding these time periods is not suitable. Therefore, from the perspective of estimation failing to incorporate these level shifts in the price series can significantly affect parameter estimation in key quantities such as the co-integration vectors. If this issue is not addressed, this could result in regular changes to portfolio allocations, resulting in additional transaction costs and other complications related to trade volumes. Therefore, in this paper we postulate that the underlying CVAR model will be a suitable model for the underlying price series in which the parameter estimation can be made less sensitive through appropriately modeling the price level shifts in the intra-day prices at open and close of markets. 

\subsection{Contribution and Structure}
The novelty of this paper involves three parts: first we develop a new matrix variate distributional model for Bayesian co-integration incorporating a mixture of matrix variate and matrix $\alpha$-stable observation errors under an error correction model (ECM) framework; the second aspect of novelty is to develop a scaled mixture of normals conjugate family of matrix variate Bayesian models for the estimation of the matrix parameters in the newly proposed model; the third aspect involves taking the non-symmetric matrix variate $\alpha$-stable setting and developing a sampling procedure for this intractable Bayesian posterior model via ABC inference. This last aspect will involve a highly non-standard combination of an adaptive MCMC matrix variate Metropolis proposal with the conjugate "symmetric" $\alpha$-stable matrix variate posterior models to obtain an efficient proposal mechanism within the ABC context. The ABC methodology will also be extended by the development of a mixed model in which aspects of the observation vector can be evaluated explicitly combined with the $\alpha$-stable random matrix observation components captured by the ABC approximation.

The multivariate $\alpha$-stable model is fitted to intra-day price level shifts over a range of currency pairs, each for 30 contract segments dating back to 1999 on minute level price data. This provides us with statistical modeling of the inter-day left shifts via generalized $\alpha$-stable models for each asset pair. We then take the parameter estimates for the $\alpha$-stable model and study the impact of naively applying the standard Johansen procedure and the Bayesian model of \cite{peters2010model} to a price series with intra-day level shifts generated from one of the more extreme currency pair $\alpha$-stable fits. This study is performed for one hundred independently generated data sets and the impact on the frequentist and Bayesian point estimators is studied. A significant impact due to the price series level shifts on the parameter estimation is observed when fitting CVAR models ignoring the price level shifts in each series. We then develop our mixture model for the noise process in the CVAR setting and we introduce two novel adaptive MCMC algorithms to work with both the simplified symmetric multivariate $\alpha$-stable model and also the more general skewed multivariate $\alpha$-stable models. Finally we conclude with a detailed data analysis both on synthetic and actual data series for pairs.

\textbf{Notation}
We denote a Gaussian random $(n \times T)$ matrix by $Y \sim N_{n,T}(\mu,\Sigma,A)$ with row dependence in $(n \times n)$ covariance matrix $\Sigma$ and column dependence in $(T \times T)$ matrix $A$. Additionally we denote the vectorization of a random matrix to a random vector by $Vec(Y)$ which will produce an $(nT \times 1)$ random vector in which the columns are successively stacked. Furthermore we denote the kronecker product or tensor product between two matrices by $\otimes$.


\section{Gaussian CVAR Model under ECM Framework}
\label{modelCVAR}
Working with the model presented in \cite{Sugita2002}, we denote the vector observation at time $t$ by $\bm{x}_{t}$. Furthermore, we assume $\bm{x}_{t}$ is an integrated of order 1, I(1), $(n \times 1)$-dimensional vector with $r$ linear cointegrating relationships. The error vector at time $t$, $\bm{\epsilon}_t$ are assumed time independent and zero mean multivariate Gaussian distributed, with covariance $\Sigma$. The Error Correction Model (ECM) representation is given by,
\begin{equation}
\label{ECM1}
\triangle \bm{x}_t = \bm{\mu} + \bm{\alpha}\bm{\beta}'\bm{x}_{t-1} + \sum_{i=1}^{p-1}\Psi_i \triangle \bm{x}_{t-i} + \bm{\epsilon}_t
\end{equation}
where $t = p, p+1, \ldots, T$ and $p$ is the number of lags. Furthermore, the matrix dimensions are: $\bm{\mu}$ and $\bm{\epsilon}_t$ are $(n \times 1)$, $\Psi_i$ and $\Sigma$ are $(n \times n)$, $\bm{\alpha}$ and $\bm{\beta}$ are $(n \times r)$. We can now re-express the model in equation (\ref{ECM1}) in a multivariate regression format, as follows
\begin{equation}
    Y = X\Gamma + Z\bm{\beta}\bm{\alpha}' + E = WB + E,
\label{eqn:1}
\end{equation}
where,
\begin{align*}
\mbox{$Y$} =
\left(
\begin{array}{cccc}
\triangle \bm{x}_{p} & \triangle \bm{x}_{p+1} & \ldots & \triangle \bm{x}_{T} \end{array}
\right) ',
\mbox{$Z$} =
\left(
\begin{array}{cccc}
\bm{x}_{p-1} & \bm{x}_{p} & \ldots & \bm{x}_{T-1} \end{array}
\right) ' \\
\mbox{$E$} =
\left(
\begin{array}{cccc}
\bm{\epsilon}_{p} & \bm{\epsilon}_{p+1} & \ldots & \bm{\epsilon}_{T} \end{array}
\right) ',
\mbox{$\Gamma$} =
\left(
\begin{array}{cccc}
\bm{\mu} & \Psi_{1} & \ldots & \Psi_{p-1} \end{array}
\right) ' \\
{\footnotesize{
\mbox{$X$} =
\left(
\begin{array}{cccc}
1 & \triangle \bm{x}_{p-1}' & \ldots & \triangle \bm{x}_{1}' \\
1 & \triangle \bm{x}_{p}' & \ldots & \triangle \bm{x}_{2}' \\
\vdots & \vdots & \ldots & \vdots \\
1 & \triangle \bm{x}_{T-1}' & \ldots & \triangle \bm{x}_{T-p+1}' \end{array}
\right)}},
\mbox{$W$} =
\left(
\begin{array}{cc}
X & Z\bm{\beta} \end{array}
\right),
\mbox{$B$} =
\left(
\begin{array}{cc}
\Gamma' & \bm{\alpha} \end{array}
\right)'
\end{align*}

Here, we let $t$ be the number of rows of $Y$, hence $t = T-p+1$, producing $X$ with dimension $t \times (1+n(p-1))$, $\Gamma$ with dimension $((1+n(p-1)) \times n)$, $W$ with dimension $t \times k$ and $B$ with dimension $(k \times n)$, where $k = 1+n(p-1)+r$. The parameters $\mu$ represents the trend coefficients, and $\Psi_{i}$ is the $i^{th}$ matrix of autoregressive coefficients and the long run multiplier matrix is given by $\Pi = \bm{\alpha}\bm{\beta}'$. 

The long run multiplier matrix is an important quantity of this model, its properties include: if $\Pi$ is a zero matrix, $\bm{x}_t$ contains n unit roots; if $\Pi$ has full rank, univariate series in $\bm{x}_t$ are (trend-)stationary; and co-integration occurs when $\Pi$ is of
rank $r<n$. The matrix $\bm{\beta}$ contains the co-integration
vectors, reflecting the stationary long run relationships between
the univariate series within $\bm{x}_t$ and the $\bm{\alpha}$ matrix contains
the adjustment parameters, specifying the speed of adjustment to
equilibria $\bm{\beta}'\bm{x}_t$.

According to \cite{Gupta99} [Theorem 2.2.1] we see that if we have a random matrix variate Gaussian $Y' \sim N_{n,T}(M,\Sigma,\Psi)$ with row dependence captured in $\Sigma$ and column dependence captured in $\Psi$, then the vectorized form, in which the columns are stacked on top of each other to make a $nT \times 1$ random vector, is multivariate Gaussian $Vec(Y) \sim N_{nT}(Vec(M),\Sigma \otimes \Psi)$. This allows us to represent the matrix variate likelihood for this regression, for the model parameters of interest $B$, $\Sigma$ and $\bm{\beta}$, by
\begin{equation}
\begin{split}
L(B,\Sigma,\bm{\beta};Y) &= (2\pi)^{-0.5nt}|\Sigma \otimes I_t|^{-0.5} \exp\left(-0.5 Vec(Y-WB)'(\Sigma^{-1}\otimes I_t^{-1})Vec(Y-WB)\right) \\
& \propto |\Sigma|^{-0.5t}\exp\left(-0.5 tr[\Sigma^{-1}(\hat{S} + R)]\right),
\end{split}
\end{equation}
where $\Sigma = Cov(\bm{\epsilon})$ and $R = (B-\hat{B})'W'W(B-\hat{B})$,
$\hat{S} = (Y-W\hat{B})'(Y-W\hat{B})$, $\hat{B} = (W'W)^{-1}W'Y$.

\section{$\alpha$-stable Models for Inter-day Differenced Price Shifts}
Noise modeling via $\alpha$-stable distributions has been suggested in several areas, such as wireless communications and in financial data analysis, see \cite{fama+r68}, \cite{godsill00}, \cite{neslehova2006infinite} and \cite{peters2010model}. $\alpha$-stable distributions possess several useful properties, including infinite mean and infinite variance, skewness and heavy tails \cite{zolotarev86} and \cite{samorodnitsky+g94}. We consider the S0 parameterization, see \cite{peters2010model} for details. Considered as generalizations of the Gaussian distribution, they are defined as the class of location-scale distributions which are closed under convolutions. Considering this class of noise process for inter-day price shifts allows us to include as a special sub-case the standard CVAR models which are assumed to have purely Gaussian innovations.

Hence, we extend the CVAR model to incorporate a composite mixture of noise processes, with $\bm{\epsilon}_t \sim \mathcal{N}(\bm{0},\Sigma)$ for intra-day samples and $\bm{\epsilon}_t \sim \mathcal{S}_{\bm{a}}(\bm{\beta},\bm{\gamma},\bm{\delta})$ for inter-day observations. In this notation, the $i$-th asset has stable iter-day error model $\epsilon^{(i)}_t \sim \mathcal{S}_{a^{(i)}}(\beta^{(i)},\gamma^{(i)},\delta^{(i)}).$ Therefore, the resulting multivariate model we consider for innovation errors $\bm{\epsilon}_t$ at time $t$ is given by dependent elements $\epsilon^{(i)}_t$,
\begin{equation}
\epsilon^{(i)}_t \sim \mathcal{N}\left(0,\sigma^{(i)}\right) \mathbb{I}\left(t \notin \bm{\tau}\right) + \mathcal{S}_{a^{(i)}}\left(\beta^{(i)}, \gamma^{(i)}, \delta^{(i)}\right) \mathbb{I}\left(t \in \bm{\tau}\right) 
\label{errorMixture}
\end{equation}
where $\mathcal{S}_{a}\left(\beta, \gamma, \delta\right) $ denotes the $\alpha$-stable distribution and $\bm{\tau}$ represents a vector of each of the first instants in time that both assets can be traded on their respective markets on each given day for the data series. 

The univariate $\alpha$-stable distribution is typically specified by four parameters: $\alpha \in (0, 2]$ determining the rate of tail decay; $\beta \in [-1, 1]$ determining the degree and sign of asymmetry (skewness); $\gamma > 0$ the scale (under some parameterizations); and $\delta \in \mathbb{R}$ the location. The parameter $\alpha$ is termed the characteristic exponent, with small and large $\alpha$ implying heavy and light tails respectively. Gaussian $(\alpha = 2, \beta = 0)$ and Cauchy $(\alpha = 1, \beta = 0)$ distributions provide the only analytically tractable sub-members of this family. In general, as $\alpha$-stable models admit no closed form expression for the density which can be evaluated point-wise (excepting Gaussian and Cauchy members), inference typically proceeds via the characteristic function, see discussions in \cite{peters2010model}. Though, intractable to evaluate point-wise, simulation of random variates is very efficient, see \cite{chambers+ms76}. This observation is crucial to the ABC based approach we develop.

We can estimate the $\alpha$-stable model parameters for the day boundary level shifts in our model in several ways, for example a quantile based generalized method of moment type procedure of \cite{McCulloch86}, or a maximum likelihood based approach of \cite{nolan97}. In addition Nolan has made available commercial and academic software for fitting univariate stable models, see his URL at \\
http://academic2.american.edu/~jpnolan/stable.html and the corresponding papers of Nolan in [Section VII] of \cite{alder+ft98} for details of the implementation.
 
The advantage of modeling the inter-day level shifts between the open and close of a market (ignoring weekends and end of segment - roll over effects) is that a statistical model of the historical behavior of these shifts, allows us to incorporate these inter-day shifts into the CVAR model which will improve the estimation of the parameters. This framework allows one to consider updating the statistical $\alpha$-stable fits sequentially over time based on the entire price series or a rolling time window based on contract lengths.

\subsection{$\alpha$-stable Empirical Assessment}
In this subsection, we first fit univariate $\alpha$-stable models to historical price series data to assess if there is evidence for modeling inter-day level shifts via an $\alpha$-stable distribution in the differenced price series. If the series indicates substantial deviation away from the standard CVAR model assumption of a Gaussian model $(\alpha=2, \beta=0)$, then a composite mixture model for the errors proposed in Equation \ref{errorMixture} becomes tenable. Otherwise, since the Gaussian distribution is also contained in the stable family, the model we propose reduces to the standard CVAR cointegration Bayesian model in \cite{peters2010model}.

Since we are analyzing the inter-day price shifts, the analysis is performed by first extracting 'daily' close/open differenced price series for each asset pairs inter-day price shifts. Daily here refers to the times when both markets for the pairs are first jointly open, or when the first market closes. Data consists of 10 minute level price data. The assets considered are AUD as Australian Dollars, CD as Canadian Dollars, FV as a US five year note, NQ as the NASDAQ mini and TU as a US two year note. In total each asset pair considers 30 contract segments, with varying numbers of days present and consecutive segment periods in time (a segment ends when a contract rolls over for one of the assets). Figure 1 shows each assets differenced price series $\triangle x_t = x_t - x_{t-1}$ from open of market each day to close of market each day, including the associated level shifts at the close/open day boundaries, for the 30 contract segments in the base currency units. We then extract these day interval differenced level shifts elements and fit them independently for each asset with a $\alpha$-stable model.
\begin{figure}[!h]
\label{fig:DIFFERENCE_SERIES}
\centering
\includegraphics[height=0.3\textheight,width=0.9\textwidth]{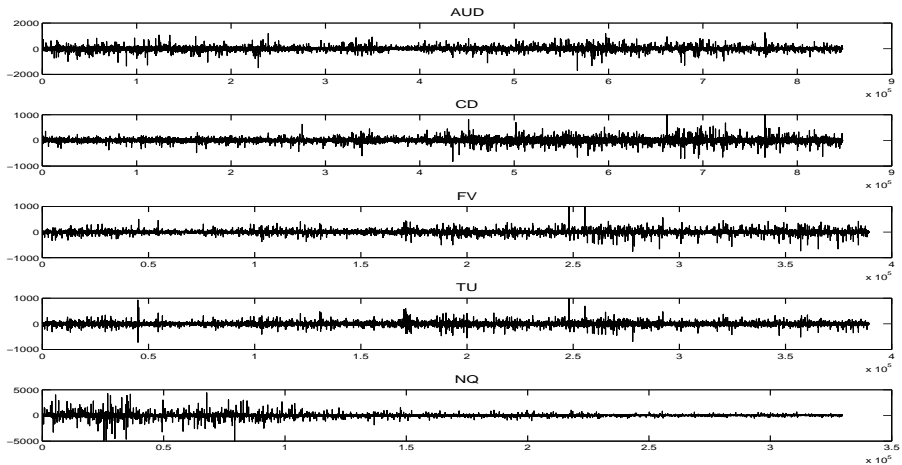}
\caption{Plots of differenced price series for the 30 contract segments.}
\end{figure}
The parameter estimation results for the $\alpha$-stable model comprised of level shift data for inter-day boundaries, in the 30 segments of each asset, are provided in Table \ref{tab:AlphaStableMLEfits}. The results are reported for the S0 parameterization for estimates obtained via Maximum likelihood procedure of \cite{nolan97}. The approach we propose here is flexible and can involve fitting the stable model to any sub segment of data required, with different stable parameter estimates per data segment. We assessed the stable fits over time by successively adding blocks of 100 days price shifts to the series and refitting the $\alpha$-stable distribution. This produces an assessment of the stability of the fitted distributions over time, we found parameter estimates to be fairly constant over the time periods considered in Table \ref{tab:AlphaStableMLEfits}.
{\small{
\begin{table}
\begin{tabular}{ccc|cccc}
\textbf{Asset i.d.} & \textbf{\# days} & \textbf{Period} & $\widehat{\alpha}$ & $\widehat{\beta}$ & $\widehat{\gamma}$ & $\widehat{\delta}$ \\ \hline 
AUD& 1535 & 05/09/99 - 30/11/05 & 1.833 (0.07) & 0.019 (0.34) & 195.365 (9) & 5.1510 (17)\\
CD & 1535 & 05/09/99 - 30/11/05 & 1.666 (0.08) & 0.028 (0.20)& 97.344 (5)& -4.699 (8)\\
FV & 960 & 05/09/99 - 18/08/03 & 1.855 (0.08) & -0.551 (0.42)& 105.134 (6)& 15.922 (11)\\
NQ & 1054 & 05/09/99 - 02/12/03 & 1.254 (0.09) & 0.009 (0.14) & 313.678 (23) & 1.673 (31)\\
TU & 960 & 05/09/99 - 18/08/03 & 1.807 (0.09) & -0.059(0.37) & 88.119 (5) & -0.088 (10)\\
\end{tabular}
\caption{Estimated Maximum Likelihood parameter estimates and in brackets the half interval 95\% confidence intervals for the estimated parameters.}
\label{tab:AlphaStableMLEfits}
\end{table}
}}

The result of this analysis suggests that it is clearly suitable to consider modeling the inter-day level shifts as distinct from a Gaussian innovations. The analysis shows that for each of the assets, the $\alpha$-stable shape parameter has 95\% confidence intervals which do not contain the Gaussian case $\alpha=2$, even with large historical data sets. Furethermore, in the case of the Canadian dollar and the Nasdaq mini index, the value of $\alpha$ obtained implies a signifcantly heavy tail model is appropriate. Additionally, several series demonstrate asymmetry, violating the assumptions of Gaussianity at these inter-day boundary points and also demonstrating that the symmetric simplification proposed in \cite{chen2010subsampling} can be invalid in many real data settings. 

\subsection{Influence of Non-Gaussian Level Shifts on CVAR Estimation}
\label{influence}
In this section we study the impact on parameter estimation for the CVAR model when failing to appropriately model the inter-day level shifts. To achieve this we consider synthetic data generated from the pair series, ($d=2$) CVAR model in Section \ref{modelCVAR} with rank $r=1$, lag $p=1$, identification constraint specified in \cite{peters2010model} and parameters specified as: $\beta = [1, -1]$; $\alpha' = [-0.002,0.001]$; $\Sigma = 100 \times \text{I}_{2\times2}$ and $\mu = [0,0]$. 

To assess the impact we generate two different groups of data series. The first consists of 100 independently generated data time series realizations of length $T=200$ using the above specified CVAR model parameters, with Gaussian innovation errors at all times, the standard CVAR model. The second consists of 100 independently generated data time series realizations, $T=200$. The difference is that the noise model is now given for each asset by
\begin{equation}
\epsilon^{(i)}_t \sim \mathcal{N}\left(0,10\right) \mathbb{I}\left(t \notin \bm{\tau}\right) + \mathcal{S}_{1.6}\left(0, 97, -4.7\right) \mathbb{I}\left(t \in \bm{\tau}\right),
\label{error}
\end{equation}
where $\alpha$-stable parameters are based on those estimated historically for the Canadian Dollar inter-day level shifts, see Table \ref{tab:AlphaStableMLEfits}. For the sake of comparison, the same Gaussian innovations are used in the rest of the time series other than those falling on a time period in which $\alpha$-stable innovation is generated.

In Figure 2 we show example comparisons of the raw price series for the model with pure Gaussian innovations (dashed line) versus the equivalently generated $\alpha$-stable mixture generated price series (solid line). In this synthetic example, we take $\bm{\tau} = \left\{t\,; \, s.t. \,\text{mod}(t,20)=0,\, \forall t \in {1,\ldots,T}\right\}$ which is equivalent to taking every 20-th noise sample from the $\alpha$-stable model fitted to this asset on historical data. 
\begin{figure}[!h]
\label{fig:StableVsGaussian}
\centering
\includegraphics[height=0.3\textheight,width=0.7\textwidth]{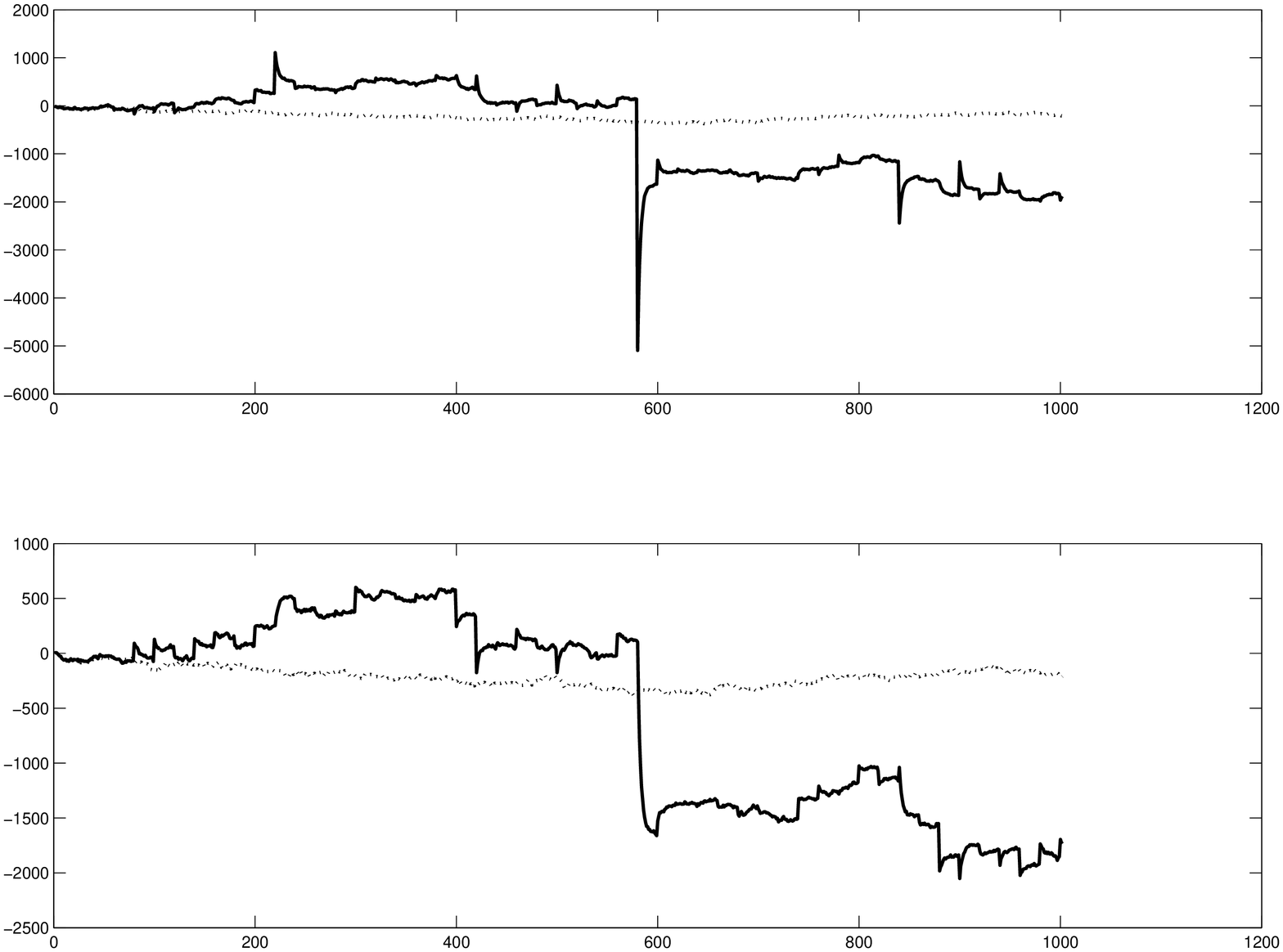}
\caption{Plots of pair raw price series for data set 1 (dashed line) standard Gaussian noise CVAR model; (solid line) Stable + Gaussian noise CVAR model.}
\end{figure}

%

For each of the synthetically generated groups of 100 data sets we estimate the parameters of the CVAR model. We compare a Maximum Likelihood based procedure, known as the Johansen procedure, see \cite{johansen1990maximum}, to a Bayesian estimation. The Bayesian CVAR model we consider utilized vague priors for all parameters, so that the likelihood would drive the parameter estimation. The posterior sampling for $\bm{\beta}$ parameters was performed via an adaptive MCMC algorithm to estimate the MMSE, as specified in \cite{peters2010model}. That is we estimated under assumed knowledge of the cointegration rank $r=1$, the nine parameters corresponding to $\Sigma$, $\bm{\alpha}, \bm{\beta}$ and $\bm{\mu}$. Both of these models estimation procedures do not account for the $\alpha$-stable noise impurity introduced, hence we can assess the impact of such noise on the parameter estimates. 

Figure 3 displays the histogram of the estimated cointegration vectors free parameter $\beta_{1,2}$ after a normalization and identification constraint, under both the Johansen procedure MLE and the Bayesian minimum mean square error (MMSE) estimate, for each data set in each group. The true parameter value used in the model to generate the data was  $\beta_{1,2}^{TRUE} = -1$. The dashed line in the figures represents the average MMSE estimate for  $\beta_{1,2}$ over the 100 data sets. The results in \textit{top sub-figures} compare the parameter estimates for the CVAR model generated with a Gaussian innovation noise (LEFT - Bayesian Estimates; RIGHT - Johansen Estimates). The Johansen procedure produced several estimates which were poor which effected the mean parameter estimate, see discussion on this point in \cite{johansen1990maximum}.
\begin{figure}[!h]
\label{fig:stabGauss}
\centering
\includegraphics[height=0.3\textheight,width=0.9\textwidth]{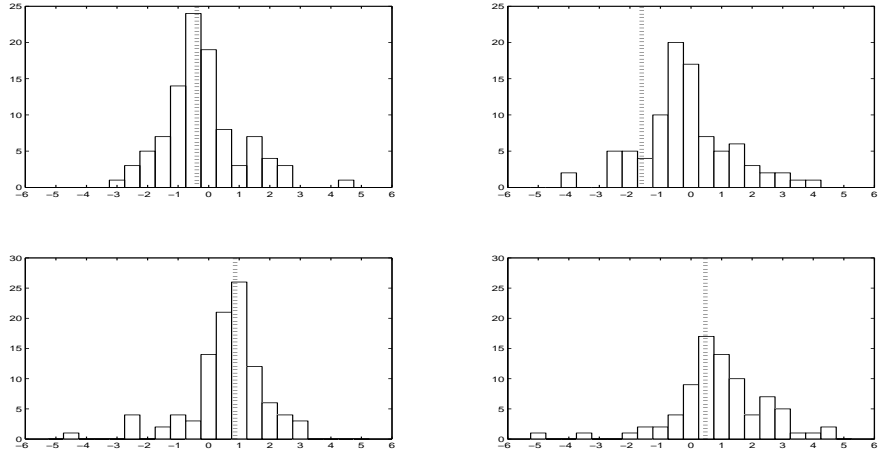}
\caption{Plots of estimated parameter $\beta_{1,2}$ for 100 data sets with and without $\alpha$-stable inter-day noise. TOP LEFT: standard Gaussian noise CVAR model estimated MMSE. TOP RIGHT standard Gaussian noise CVAR model Johansen MLE.
BOTTOM LEFT: $\alpha$-stable + Gaussian noise model estimated MMSE (ignoring stable noise presence). BOTTOM RIGHT: $\alpha$-stable + Gaussian noise model Johansen MLE (ignoring stable noise presence).}
\end{figure}

\vspace{0.1cm}
The results demonstrate that both the Johansen and the Bayesian estimates for the cointegration vector $\bm{\beta}$ in each of the 100 data sets are severely affected by the presence of the inter-day jumps, modeled here by the $\alpha$-stable noise. Therefore to avoid bias in the parameter estimates obtained in the CVAR model, one must appropriately model the inter-day level shifts in the price series.

\section{ABC Bayesian CVAR Models }
\label{BaysianCVAR_LHF}
Here we extend the class of Bayesian CVAR models presented in \cite{peters2010model} and \cite{sugita2009monte} to include the composite $\alpha$-stable noise model developed. This will allow us to then formulate a Bayesian estimation procedure for the parameters in this model. In doing so we are able to estimate the parameters of the CVAR model with out the bias introduced by inappropriate model assumptions as assessed in Section \ref{influence}. Note that due to the fact that the general $\alpha$-stable model does not admit a tractable density, this directly impacts on the ability to apply the standard Johansen procedure, as the likelihood can no longer be evaluated point-wise. Alternatives in such cases include indirect inference, see \cite{gourieroux1993indirect}. This would generalize the symmetric simplification proposed in \cite{chen2010subsampling}.   

Instead we formulate a novel ABC or approximate Bayesian computation (ABC) solution. ABC Bayesian modeling is a new class of statistical techniques specifically designed for modeling when the likelihood and thus the posterior distribution is intractable. These have now been studied and applied in a range of settings, see \cite{peters2010chain} and \cite{peters2006bayesian} for ABC modeling for financial risk and insurance contexts. In addition, there are now several methodological papers and reviews available for this new class of modeling technique, see \cite{peters+fs08}, \cite{tavare+bgd97}, \cite{fearnhead2010semi}, \cite{beaumont2009adaptivity} and the review of \cite{sisson-likelihood}.

We make identical model assumptions and restrictions for the Bayesian CVAR model as in \cite{peters2010model}. In particular, for any non-singular matrix $A$, the matrix of long run multipliers $\Pi = \bm{\alpha}\bm{\beta}'$ is indistinguishable from $\Pi = \bm{\alpha} A A^{-1} \bm{\beta}'$, see \cite{koop2006bayesian}. We remove this problem by incorporating a non unique identification constraint by imposing $r^2$ restrictions as follows $\bm{\beta} = [I_{r}, \bm{\beta}_{*}']'$, where $I_{r}$ denotes the $r \times r$ identity matrix, see \cite{kleibergen2009shape}. We first specify our prior structure and then separate the problem into two sub cases, the symmetric $\alpha$-stable case and the general skewed $\alpha$-stable model. We present the Bayesian model for estimation of $\bm{\beta}$, $B$ and $\Sigma$ conditional on the rank $r$ under each of these settings. 
\subsection{Prior}
\label{Priors}
The prior model is identical to the choice of \cite{peters2010model} and \cite{Sugita2002}, which produces conjugate posterior distributions for matrix variate parameters $\Sigma$ and $B$. In the new composite noise model we develop we must re-derive the Bayesian models in the presence of the $\alpha$-stable inter-day noise. In general conjugacy is lost for the general asymmetric noise models in Equation \ref{errorMixture}. However, we derive a novel conjugacy under transformation in the symmetric case via a scaled mixture of Normals (SMiN) representation of the $\alpha$-stable inter-day model. 

\begin{itemize}
    \item $\bm{\beta}' \sim  N(\bar{\bm{\beta}}',Q \otimes H^{-1})$ where $N(\bar{\bm{\beta}},Q \otimes H^{-1})$ is the matrix-variate Gaussian distribution with prior mean $\bar{\bm{\beta}}$, Q  is a $(r \times r)$ positive definite matrix, H a $(n \times n)$ matrix.
    \item $\Sigma \sim IW(S,h)$ where $IW(S,h)$ is the Inverse Wishart distribution with h degrees of freedom and S is an $(n \times n)$ positive  definite matrix.
    \item $B'|\Sigma \sim N(P',\Sigma \otimes A^{-1} )$ where $N(P,\Sigma \otimes A^{-1})$ is the matrix-variate Gaussian distribution with h degrees of freedom and S is an $(n \times n)$ positive definite matrix.
\end{itemize}

\subsection{Derivation of a Conjugate Matrix-variate SMiN Bayesian CVAR Model}
In this section a novel matrix variate Bayesian conjugate model is derived for the mixture of noise processes in the ECM framework. Lemma 1 and Lemma 2 combined with Theorem 1 demonstrate that under a specifically designed transformation of the vectorized matrix of observations, we can obtain a joint likelihood for the $\alpha$-stable and Guassian innovations mixture model in the un-vectorized matrix variate observations which is matrix-variate Gaussian with explicit covariance matrix under the transformation. This will be critical as we wish to obtain a Bayesian conjugate model for the posterior matrix parameters. In addition to the covariance structure, Lemma 3 and Theorem 2 then derive the form of the mean matrix for this matrix variate likelihood, via a well known tensor product identity on vectorized transformed data. To achieve this we consider a special form of non-negative tensor factorization of our transformation matrix. In addition we prove that the solution to the mean struture parameter matrix in the transformed model, can be uniquely recovered under the transformation developed, given estimates of the transformed parameters. Therefore we can sample the transformed posterior distribution and then invert posterior samples via the transformation to obtain un-transformed samples uniquely. Finally, Theorem 3 derives the conjugate model for the matrix variate parameters of the posterior under the transformation. This is meaningful as it allows us to exploit existing results developed for matrix variate distributions in the CVAR ECM framework. The other important resuls is that this allows us to reduce the posterior dimension significantly, as we do not need to parameterize the posterior covariance matrix for the vectorized observations which would be dimenson $nT \times nT$, instead allowing us to work with an $n \times n$ posterior matrix. Clearly, a significant dimension reduction, especially in the setting of financial data, where the number of data points $T >> n$ is of the order of 100's to 1,000's. 

When the noise model in Equation \ref{errorMixture} is strictly symmetric, ie. the $\alpha$-stable inter-day noise model is symmetric, it admits an exact SMiN representation, see \cite{godsill00}. This involves models of the form 
\begin{equation}
\begin{split}
\epsilon^{(i)}_t 
&\sim \mathcal{N}\left(0,\sigma^{(i)}\right) \mathbb{I}\left(t \notin \bm{\tau}\right) + \mathcal{N}\left(\delta^{(i)}, \gamma^{(i)}\lambda^{(i)}\right) \mathbb{I}\left(t \in \bm{\tau}\right),
\end{split}
\label{errorMixture}
\end{equation}
with auxiliary scale variables distributed as $\lambda^{(i)} \sim \mathcal{S}_{a^{(i)}/2}\left(0,1,1\right)$.

For simplicity we assume a lag $p=1$, this can be extended trivially under our framework. We will first take all the vector observations for times 1 to T (mixed sets of (inter-day) SMiN and standard (intra-day) CVAR Gaussian innovation noise random vector observations), denoted $\bm{y}_{1:T}$ with dimension $T \times n$, which will have a log-likelihood model given by {\footnotesize{
\begin{equation}
\begin{split}
\mathcal{L}&(\Sigma, B, \bm{\beta},\bm{\lambda},\bm{\alpha},\bm{\gamma},\bm{\delta};\bm{y}_{1:T}) \\
 &= \log\left((2\pi)^{-0.5n\tilde{t}}|\Sigma \otimes I_{\tilde{t}}|^{-0.5} \exp\left(-0.5 Vec(\tilde{Y}-\tilde{W}B)'(\Sigma^{-1}\otimes I_{\tilde{t}}^{-1})Vec(\tilde{Y}-\tilde{W}B)\right)\right) \mathbb{I}\left(t \notin \bm{\tau}\right)\\
&+ \log\left((2\pi)^{-0.5n(T-\tilde{t})}|\bm{D_{\lambda}} \otimes I_{(T-\tilde{t})}|^{-0.5} \exp\left(-0.5 Vec(\chi-W_{(T-\tilde{t})}B)'\left(\left(\bm{D_{\lambda}}\right)^{-1}\otimes I_{(T-\tilde{t})}^{-1}\right)Vec((\chi-W_{(T-\tilde{t})}B))\right)\right)\mathbb{I}\left(t \in \bm{\tau}\right)\\
\end{split}
\label{LogLHModel}
\end{equation}
}}
where $\tilde{Y}$ represents the matrix of observation differenced price vectors corresponding to intra-day prices with a total of $\tilde{t}$ rows and $\tilde{W}$ is the corresponding matrix for $\tilde{Y}$. In addition $\chi = Y_{-\tilde{Y}}-\bm{1}_{k}\bm{\delta}^T$ corresponds to the inter-day observation matrix of observation differenced price vectors not including rows for $\tilde{Y}$ after subtracting of the location parameters for each $\alpha$-stable fit, given by $\bm{\delta} = \left[\delta^{(1)},\ldots,\delta^{(n)}\right]'$. The definition of $W_{(T-\tilde{t})}$ is the matrix for $W$ corresponding to the observation vectors taken from the set of intra-day times when $t \in \bm{\tau}$.
The vectors $\bm{\lambda} = \left[\tilde{\lambda}^{1}\gamma^{1},\ldots,\tilde{\lambda}^{n}\gamma^{n}\right]$ are the scale parameters in the SMiN representation and $\bm{D_{\lambda}}$ is a diagonal matrix with each value of $\bm{\lambda}$ in the diagonal. This vectorized representation is instructive to understand the model, however to exploit conjugacy present it will be beneficial to re-represent the likelihood in a matrix variate decomposed form specified in Lemma 1. \\

{\large \textbf{Lemma 1.}} 
\textsl{Utilizing the assumption of conditional independence of the observation vectors given model parameter matrices and vectors $\Sigma, B, \bm{\beta},\bm{\lambda},\bm{\alpha},\bm{\gamma},\bm{\delta}$ which states $E\left[\bm{y}_s,\bm{y}_t\right]=E\left[\bm{y}_s\right]E\left[\bm{y}_t\right] \: \forall s,t \, s \neq t$ and additionally the results from \cite{Gupta99} ([Theorem 2.2.1], [Theorem 2.3.11]) and the trace identity and determinant identities of \cite{Gupta99} [Theorem 1.2.21 (v and x)] we can specify the complete grouped vectorized likelihood. That is we consider a reordered version of $Y = \bm{y}_{1:T}$ denoted generically by $Vec(Y_*) \sim N_{nT}\left(Vec(M_*),\Sigma_* \otimes \Psi_* \right)$ which gives the grouped likelihood model:
{\small{
\begin{equation}
\begin{split}
L(\Sigma, & B, \bm{\beta},\bm{\lambda},\bm{\alpha},\bm{\gamma},\bm{\delta};Y_*) \\
&= (2\pi)^{-0.5nT}|\Sigma_* \otimes \Psi_{*}|^{-0.5} \exp\left(-0.5 Vec(Y_*-M_*)'(\Sigma_*^{-1}\otimes \Psi_{*}^{-1})Vec(Y_*-M_*)\right) \\
&\propto |\Sigma_*|^{-0.5T}|\Psi_*|^{-0.5n}\exp\left(-0.5 \text{tr}\left\{ \Sigma_*^{-1}(Y_*-D_*-W_*B)'\Psi_{*}^{-1}(Y_*-_*-W_*B)\right\}\right)
\end{split}
\label{eqn:MatrixLH}
\end{equation}
}}where we have ordered the observation vectors $$Y_* = \bm{y}_{1:T} = [\bm{y}_{1} \, \bm{y}_{2} \,\ldots \,\bm{y}_{\tau_1-1} \,\bm{y}_{\tau_1+1} \,\ldots \,\bm{y}_{T} \, \bm{y}_{\tau_1}\, \bm{y}_{\tau_2} \,\ldots \,\bm{y}_{\tau_{i_D}}]'$$ and there are a total of $i_D$ inter-day boundaries in the series. In addition we define the appropriate likelihood matrices as follows for a general covariance matrix structure $\Sigma_* \otimes \Psi_{*}$ (for $\Sigma_*$ a $n \times n$ matrix and $\Psi_{*}$ a $T \times T$ matrix),  
\begin{align*}
\mbox{$D_*$} =
\left(
\begin{array}{c}
\bm{0} \\
\bm{1}_{i_D}\bm{\delta}^T \\
\end{array}
\right),
\mbox{$W_*$} =
\left(
\begin{array}{c}
\tilde{W} \\
W_{(T-\tilde{t})} \\
\end{array}
\right).
\end{align*}
}\\
We now present some remarks about grouping all observations from intra-day and inter-day into a single matrix-variate Gaussian likelihood model.\\

\textbf{Remark 2:} \textsl{Lemma 1 states that generically the observations can be reordered to form the $(n \times T)$ Gaussian random matrix $Y_*$ with the first $\tilde{t}$ columns corresponding to the \textbf{intra-day} price differences and the remaining $T - \tilde{t}$ columns from the SMiN observations. In addition we can represent the matrix variate Gaussian as having a covariance structure given generically by $\Sigma_* \otimes \Psi_{*}$, where $\Sigma_*$ corresponds to the row dependence and $\Psi_*$ captures the column dependence. Lemma 1 also presented the required mean structure for this combined matrix-variate likelihood.}\\

\textbf{Remark 3:} \textsl{To relate the matrix variate Gaussian model, obtained from Lemma 1, to the original likelihood model in \ref{LogLHModel} we need to find a relationship to identify the sufficient statistics matrices, $\Sigma_*$ and $\Psi_*$ with the original likelihood model. Under this reordered and repacked matrix variate Gaussian, the independent columns of the random matrix is no longer true, that is $\Psi_*$ is only diagonal when $\bm{D_{\lambda}} = \Sigma$.}\\

\textbf{Remark 4:} \textsl{Maintaining the conjugacy structures developed in \cite{peters2010model} [Section 3] and \cite{Sugita2002}, for the standard matrix variate Gaussian noise Bayesian CVAR model is beneficial for inference and sampling. This would require us to identify the sufficient statistics, $(M_*,\Sigma_*,\Psi_*)$, for the grouped matrix variate Gaussian model in Lemma 1, and to have $\Sigma_*=\Sigma$ and $\Psi_*$ diagonal, as this will preserve conjugacy results, conditional on parameters from the fitted $\alpha$-stable SMiN intra-day noise model. This would allow us to specify a matrix variate prior only on a matrix $\Sigma_*$ which is $n \times n$ rather than a multivariate covariance which is $nT \times nT$ thus providing a significant dimension reduction in our posterior model parameters to be estimated.}\\

Lemmas 2, 3 and Theorems 1, 2 and 3 allow us to identify the sufficient statistics and then transform the vectorized random observation matrix $Y_*$ to recover the desired conjugacy properties discussed in Remarks 2, 3 and 4.\\

{\large \textbf{Lemma 2.}} \textsl{Using [Definition 2.2.1] and [Theorem 2.2.1] of \cite{Gupta99}, the random vector $Vec(Y_*)$ is conditionally a multivariate Gaussian random vector of dimension $nT \times 1$. Using Lemma 1 and the SMiN CVAR model assumption of conditional independence, but not identically distributed, Gaussian observation random vectors we can explicitly identify the mean and covariance structure of the vectorized observation matrix $Vec(Y_*)$ in terms of the original CVAR model matrices as follows,  
$$Cov(Vec(Y_*)) = \Sigma_*\otimes\Psi_*=\left(
\begin{array}{cc}
  \Sigma \otimes I_{\tilde{t}} & \bm{0} \\
 \bm{0} &  \bm{D_{\lambda}} \otimes I_{(T-\tilde{t})}\\
\end{array}
\right).$$
In addition we can obtain the covariance of $Vec(Y_*')$ as
$$Cov(Vec(Y_*')) = \Psi_*\otimes\Sigma_*=\left(
\begin{array}{cc}
   I_{\tilde{t}} \otimes \Sigma & \bm{0} \\
 \bm{0} &   I_{(T-\tilde{t})} \otimes \bm{D_{\lambda}}\\
\end{array}
\right).$$
}
Having identified the covariance structure for the vectorized reordered observation matrix, we present Theorem 1 to address Remark 4 which pertains to maintaining a likelihood structure that will admit conjugacy under the priors presented in Section \ref{Priors}.\\
{\large \textbf{Theorem 1.}} \textsl{Given Lemma 1 and Lemma 2 which provide us with a $(nT \times 1)$ random vector $Vec(Y_*)$ conditionally distributed according to a multi-variate Gaussian distribution, under a transformation by a $nT \times nT$ matrix $Q_*$ we can obtain a transformed random vector denoted $Vec(Z_*) = Q_*Vec(Y_*)$ which is also multivariate Gaussian. Using \cite{lutkepohl2005new} [Proposition B.2] we obtain, for \\ $Vec(Y_*) \sim N\left(Vec(M_*),\Sigma_* \otimes \Psi_* \right)$, a transformed random vector \\$Vec(Z_*) = Q_*Vec(Y_*) \sim N(Q_*Vec(M_*),Q_*^T(\Sigma_* \otimes \Psi_*)Q_*)$. \\If we select the transformation 
$$Q_* = \left(
\begin{array}{cc}
I_{n} \otimes I_{\tilde{t}} & \bm{0} \\
 \bm{0} &  Q \otimes I_{(T-\tilde{t})}\\
\end{array}
\right)$$
then we can obtain a particular form for the $n \times T$ un-vectorized random matrix for $Z_*$ which has a covariance structure based on the original covariance for the Gaussian inter-day innovation noise $\Sigma$. That is we obtain $Z_* \sim N_{n,T}(\mu_*,\Sigma,I_T)$. In addition we can define 
$$Q_{*t} = \left(
\begin{array}{cc}
 I_{\tilde{t}} \otimes I_{n} & \bm{0} \\
 \bm{0} &  I_{(T-\tilde{t})} \otimes Q\\
\end{array}
\right)$$
such that when it is used to transform $Q_{*t}Vec(Y_*')$ we obtain $Q_{*t}Vec(Y_*') \sim N_{n,T}(\mu_*,I_T,\Sigma)$ and we also have that 
$Z_*' = Q_{*t}Vec(Y_*')$.
}\\
\textbf{Proof:} To prove the covariance structure of the transformed random vector under this particular transformation has this special tensor product factorization we consider the new covariance structure for $Vec(Z_*)$ which will be given by
\begin{equation*}
\begin{split}
\mathbb{C}ov(Vec(Z_*)) =
&\left(
\begin{array}{cc}
I_{n} \otimes I_{\tilde{t}} & \bm{0} \\
 \bm{0} &  Q \otimes I_{(T-\tilde{t})}\\
\end{array}
\right)^T
\left(
\begin{array}{cc}
\Sigma \otimes I_{\tilde{t}} & \bm{0} \\
 \bm{0} &  \bm{D_{\lambda}} \otimes I_{(T-\tilde{t})}\\
\end{array}
\right)
\left(
\begin{array}{cc}
I_{n} \otimes I_{\tilde{t}} & \bm{0} \\
 \bm{0} &  Q \otimes I_{(T-\tilde{t})}\\
\end{array}
\right) \\
&=\left(
\begin{array}{cc}
\Sigma \otimes I_{\tilde{t}} & \bm{0} \\
 \bm{0} & \left(Q^T\bm{D_{\lambda}}Q \otimes I_{(T-\tilde{t})}\right)\\
\end{array}
\right)
\end{split}
\end{equation*}
We can therefore obtain $\mathbb{C}ov(Vec(Z_*)) = \Sigma \otimes I_T$ by solving the equation $Q^T\bm{D_{\lambda}} Q = \Sigma $ for matrix $Q$. We can make use of the fact that the $n \times n$ matrix $\bm{D_{\lambda}}$ is diagonal and the covariance matrix $\Sigma$ is real and symmetric with an eigen decomposition $\Sigma = VFV^T$ with diagonal eigen values matrix $F$. Therefore if we select $Q=S^{\frac{1}{2}}U^T$ where $S^{\frac{1}{2}}$ is the diagonal matrix with the elements $S_{ii}= \sqrt{\frac{F_{ii}}{\bm{D_{\lambda,ii}}}}$ then the matrix $U$ is the orthonormal matrix of eigen vectors for $\Sigma$, that is $U=V$. The proof for the transformation $Q_{*t}$ of $Vec(Y_*')$ follows trivially from this result. $\Box$\\

Hence, we have transformed the observation vector $Vec(Y_*)$ via matrix $Q_*$ to obtain a new random vector which when un-vectorized produces a matrix variate Gaussian with row dependence given by $\Sigma$ and column dependence given by $I_T$.  This therefore recovers the conditional independence property of each vector observation whilst identifying under the transformation the identity $\Sigma_* = \Sigma$ and $\Psi_* = I_T$. Therefore the matrix variate likelihood for transformed observations $\bm{z}_{1:T}$ is given by Lemma 3. \\

{\large \textbf{Lemma 3.}} \textsl{ Under [Definition 2.2.1] and [Theorem 2.2.1] of \cite{Gupta99}, the likelihood of the transformed observations is given by
{\small{
\begin{equation*}
\begin{split}
L&(\Sigma, B, \bm{\beta},\bm{\lambda},\bm{\alpha},\bm{\gamma},\bm{\delta}, Q; \bm{z}_{1:T}) \\
&\propto |\Sigma_* \otimes I_{T}|^{-0.5} \exp\left(-0.5 \left(Vec(Z_*)-Q_*Vec(D_*-W_*B)\right)'(\Sigma_*^{-1}\otimes I_{T}^{-1})\left(Vec(Z_*)-Q_*Vec(D_*-W_*B)\right)\right)\\
\end{split}
\end{equation*}
}}
Then applying the identity in [Theorem 1.2.22] of \cite{Gupta99}, given by 
\begin{equation}
\label{VecMean}
\left(B' \otimes A \right)Vec(X) = Vec(AXB),
\end{equation}
we can rearrange the mean structure of the likelihood model. We can make an arbitrary choice of factorization of $Q_*$ into the form $Q_* = G \otimes H$ with the only constraints that $G$ is $(p \times n)$ and that $H$ is $(q \times T)$ dimensions, with $pq = nT$. There are several solutions to this class of tensor factorization, we will present our factorization in Lemma 5.
Hence, we rearrange the mean structure in the likelihood as,
{\small{
\begin{equation}
\begin{split}
L&(\Sigma, B, \bm{\beta},\bm{\lambda},\bm{\alpha},\bm{\gamma},\bm{\delta}, Q; \bm{z}_{1:T}) \\
&\propto |\Sigma_* \otimes I_{T}|^{-0.5} \exp\left(-0.5 \left(Vec(Z_*)-Q_*Vec(D_*-W_*B)\right)'(\Sigma_*^{-1}\otimes I_{T}^{-1})\left(Vec(Z_*)-Q_*Vec(D_*-W_*B)\right)\right)\\
&\propto |\Sigma_* \otimes I_{T}|^{-0.5} \exp\left(-0.5 \left(Vec(Z_*)- Vec(\tilde{D}_*-\tilde{W}_*\tilde{B})\right)'(\Sigma_*^{-1}\otimes I_{T}^{-1})\left(Vec(\tilde{D}_*-\tilde{W}_*\tilde{B})\right)\right)
\end{split}
\end{equation}
}}
defining $\tilde{D}_* = HD_*G^T$, $\tilde{W}_*=HW_*$ and $\tilde{B}=BG^T$. This then allows us to re-express the likelihood model in the form
{\small{
\begin{equation}
\begin{split}
p&( \bm{z}_{1:T}|\Sigma, B, \bm{\beta},\bm{\lambda},\bm{\alpha},\bm{\gamma},\bm{\delta}, Q) \\
&\propto |\Sigma_*|^{-0.5T}|I_T|^{-0.5n}\exp\left(-\frac{1}{2} \text{tr}\left\{ \Sigma_*^{-1}(Z_*-\tilde{D}_*-\tilde{W}_*\tilde{B})'(Z_*-\tilde{D}_*-\tilde{W}_*\tilde{B})\right\}\right)\\
&= |\Sigma_*|^{-0.5T}\exp\left(-\frac{1}{2} \text{tr}\left\{ \Sigma_*^{-1}\left(\widehat{\tilde{S}}_* + (\tilde{B} - \widehat{\tilde{B}}_*)'\tilde{W}_*'\tilde{W}_*(\tilde{B} - \widehat{\tilde{B}}_*)\right)\right\}\right)\\
\end{split}
\end{equation}
}}
with $\widehat{\tilde{B}}_* = \left(\tilde{W}_*'\tilde{W}_*\right)^{-1}\tilde{W}_*'(Z_*-\tilde{D}_*)$ and $\widehat{\tilde{S}}_* = \left(Z_* - \tilde{D}_* - \tilde{W}_*\widehat{\tilde{B}}_*\right)'\left(Z_* - \tilde{D}_* - \tilde{W}_*\widehat{\tilde{B}}_*\right)$.
If lags of $p > 1$ are of interest, this approach can still be used, but the block diagonal covariance matrix will involve more sub-blocks.
}\\
We can now comment on the possible solutions to this tensor factorization.

\textbf{Remark 5:} \textsl{Typically the basic Singular Value Decomposition is applied to perform a tensor factorization - but this will be difficult in our setting as we are required to enforce the sub-matrix constraints that the first factored matrix must be $(p \times n)$ with $n$ columns and the second $q \times T$ with T columns. Another solution would be to search over all subspaces for the $p$ and $q$ combinations to satisfy $pq = nT$ for a set of matrices that minimizes a matrix norm. There is a rich literature on such tensor factorizations and the interested reader is referred to numerical algorithms for rank-k tensor approximations which generalize the SVD such as the orthogonal tensor decompositions (Higher-Order SVD) of \cite{shashua2005non}, \cite{de2004dimensionality} or 3-way decompositions of \cite{harshman1970foundations} known as PARAFAC and the Non-Negative Tensor Factorization (NTF) in \cite{friedlandera2008computing}.}\\

In Theorem 2 we provide a specific tensor factorization to satisfy Lemma 3. It is important to obtain a specific factorization which allows us to decompose the transformation matrix into a tensor factorization which admits at least one solution to the original mean structure for $B$. When multiple solutions are present we can handle this in our Bayesian framework through imposing constraints post sampling, as typically performed in these situations in which there are complications with identifiability, see \cite{celeux2000computational}. We can provide a unique solution for the original mean structure for $B'$ given $\tilde{B}'$.\\

{\large \textbf{Theorem 2.}} \textsl{Given transformed observations, $Z_*'$, we obtain an analytic tensor factorization for the transformation matrix $Q_{*t}$ satisfying the dimensionality constraints on the tensor factors in Lemma 3, given by
\begin{equation*}
\begin{split}
Q_{*t} = \sum_{i=1}^T \sum_{j=1}^T U_{ij} \otimes Q_{i,j}
\end{split}
\end{equation*}
where $Q_{i,j}$ represents the $(i,j)$-th sub-block of dimension $n \times n$ in the $nT \times nT$ transform matrix $Q_{*t}$ and $U_{ij}$ represents the $(T \times T)$ matrix whose $ij$-th element is 1 and whose remaining elements are 0. This particular choice of factorization ensures that a unique solution to $B'$ is attainable given $\tilde{B}$. This will be particularly important for the conjugate Bayesian model in Theorem 3. The mean structure under the transformation is given by,
\begin{equation*}
\begin{split}
\mathbb{E}\left[Vec(Z_*')\right] &= Q_{*t}Vec(D_*'-B'W_*')\\
&= \sum_{i=1}^T Vec(Q_{ii}(D_*'-B'W_*')U_{ii}')
\end{split}
\end{equation*}
This allows us to make explicit the mean structure of the matrix variate transformed data likelihood of Lemma 3 by identifying the following elements 
$\tilde{D}_*' = \sum_{i=1}^T Q_{ii}D_*'U_{ii}'$,
$\tilde{W}_*' = \sum_{i=1}^T W_*' U_{ii}$ and
$\tilde{B}' = \sum_{i=1}^T Q_{ii} B'$.}\\

\textbf{Proof:} Using the identity [(1.29) p. 343] of \cite{harville2008matrix} we can exploit the fact that the transformation matrix $Q_{*t}$ we have selected is a square $nT \times nT$ matrix which has a $n \times n$ block diagonal structure. Hence we will consider the following structure in $Q_{*t}$
\begin{equation*}
\begin{split}
&\left(
\begin{array}{cccc}
Q_{11} & Q_{12} & \cdots & Q_{1T} \\
\vdots & \vdots &  & \vdots \\
Q_{T1} & Q_{T2} & \cdots & Q_{TT} \\
\end{array}
\right)^T
\end{split}
\end{equation*}
with each sub matrix $Q_{ij}$ being selected as $(n \times n)$ matrix. We can then obtain the following tensor factorization, using the fact that all $Q_{i,j}$ matrices will be comprised of $0$ elements other than those with $i=j$ giving a sparse representation
\begin{equation*}
\begin{split}
Q_{*t} &= \sum_{i=1}^T \sum_{j=1}^T U_{ij} \otimes Q_{ij} = \sum_{j=1}^T U_{ii} \otimes Q_{ii}.
\end{split}
\end{equation*}
As above, $U_{ij}$ represents the $(T \times T)$ matrix whose $ij$-th element is 1 and whose remaining elements are 0 and we have used the fact that we have specifically selected the transformation matrix $Q_{*t}$ as $n \times n$ block diagonal. Under this factorization the mean structure we obtain in the likelihood model in Theorem 1 with application of the identity in [Theorem 1.2.22] of \cite{Gupta99} shown in Equation \ref{VecMean}, is given by
\begin{equation*}
\begin{split}
\mathbb{E}\left[Q_{*t}Vec(Y_*')\right] &= Q_{*t}Vec(D_*'-B'W_*')\\
&= \sum_{i=1}^T \sum_{j=1}^T \left(U_{ij} \otimes Q_{ij}\right)Vec(D_*'-B'W_*')\\
&= \sum_{i=1}^T Vec(Q_{ii}D_*'U_{ii}'-Q_{ii}B'W_*'U_{ii}'))\\
\end{split}
\end{equation*}
This allows us to make explicit the mean structure of the matrix variate transformed data likelihood by identifying the following elements 
$\tilde{D}_*' = \sum_{i=1}^T Q_{ii}D_*'U_{ii}'$,
$\tilde{W}_*' = \sum_{i=1}^T W_*'U_{ii}'$ and
$\tilde{B}' = \sum_{i=1}^T Q{ii} B' $.\\
Finally, we note that we can uniquely solve the system 
$$\tilde{B}' = \sum_{i=1}^T Q_{ii} B' $$
for $B'$ given $\tilde{B}'$. This is due to the fact that the matrices $Q_{ii}$ for $i < T$ are constructed from identity matrices and the case of $i=T$ is constructed in our transform as a real matrix of eigen vectors of covariance matrix $\Sigma$, which is therefore invertible. We can therefore obtain the unique solution for $B'$ as
$$B' = \tilde{B}' \left((T-1)I_{n} + Q_{TT}\right)^{-1}.$$
$\Box$\\
Under the transformed observation vector we utilize an identical prior model for the transformed mean structure as specified in Section \ref{Priors} to obtain conjugacy for the transformed prior-likelihood model.\\

{\large \textbf{Theorem 3.}} \textsl{Under Theorem 1, Theorem 2, Lemma 1, 2 and 3 and conditional on parameter estimates of the multivariate $\alpha$-stable statistical model, $\mathcal{S}_{\bm{\alpha}}\left(\bm{\beta},\bm{\gamma},\bm{\delta}\right)$, fitted to historical price series inter-day level shifts for each asset in the CVAR model, the following posterior conjugacy properties are satisfied for the prior choices in Section 
\ref{Priors}:
\begin{enumerate}
\item[\textbf{Conditional 1:} ]{Conditional on the re-arranged un-transformed subset of observation vectors from intra-day prices matrix $\widetilde{Y}$ we obtain an Inverse Wishart distribution for $$p(\Sigma|\bm{\beta},\bm{\lambda},\bm{\alpha},\bm{\gamma},\bm{\delta},\widetilde{Y}) \propto |S_{\tilde{Y}}|^{(t+h)/2}|\Sigma|^{-(t+h+n+1)/2}\exp\left(-0.5 tr(\Sigma^{-1}S_{\tilde{Y}})\right);$$
where $S_{\tilde{Y}}$ is defined to be given by
$$S_{\tilde{Y}} = S + \widehat{S} + (P-\widehat{B})'\left[A^{-1} + (W'W)^{-1}\right]^{-1}\left(P - \widehat{B}\right).$$
}
\item[\textbf{Conditional 2:} ]{Under the SMIN model and conditional on the re-arranged transformed complete vector of observations for intra and inter-days, $Vec(Z_*) = Q_* Vec(Y_*)$ we obtain a Matrix-variate Gaussian for $$p(\tilde{B}|\bm{\beta},\bm{\lambda},\bm{\alpha},\bm{\gamma},\bm{\delta},\Sigma,Z_*,Q_*) \propto |A_{Z_*}|^{n/2}|\Sigma|^{-k/2}\exp\left(-0.5 tr\left(\Sigma^{-1}(\tilde{B}-B_{Z_*})'A_{\star}(\tilde{B}-B_{Z_*})\right)\right)$$
where $A_{Z_*} = \tilde{A} + \tilde{W}_*'\tilde{W}_*$ and $B_{Z_*} = \left(\tilde{A} + \tilde{W}_*'\tilde{W}_*\right)^{-1}\left(\tilde{A}\tilde{P} + \tilde{W}_*'\tilde{W}_*\widehat{\tilde{B}}_*\right)$.
}
\item[\textbf{Conditional 3:} ]{Under the SMIN model and conditional on the re-arranged transformed complete vector of observations for intra and inter-days, $Vec(Z_*) = Q_* Vec(Y_*)$ we obtain the marginal matrix-variate posterior for the cointegration vectors, $\bm{\beta}$ given by
$$p(\bm{\beta}|\bm{\lambda},\bm{\alpha},\bm{\gamma},\bm{\delta},Z_*,Q_*) \propto p(\bm{\beta})|S_{Z_*}|^{-(t+h+1)/2}|A_{Z_*}|^{-n/2}.$$
for 
$$S_{Z_*} = S + \widehat{\tilde{S}}_* + (P-\widehat{\tilde{B}}_*)'\left[\tilde{A}^{-1} + (\tilde{W}_*'\tilde{W}_*)^{-1}\right]^{-1}\left(P - \widehat{\tilde{B}}_*\right)$$
and $A_{Z_*}$ defined in Conditional 2.
}
\item[\textbf{Conditional 4:} ]{Under the SMIN model we obtain a the marginal distribution for each random variable $\lambda^{i}$ in the $n \times 1$ random vector $\bm{\lambda}$ given by
$$p(\lambda_i|\bm{\alpha},\bm{\gamma},\bm{\delta},\chi,\tilde{\bm{B}},Q_*,\bm{\beta}) \propto \prod_{t \in \tau} \mathcal{N} \left( \epsilon_t^{i} ;0,\lambda_i \gamma_i \right) \times S_{a_i/2} \left( \lambda_i ;0,1,1 \right)$$
where for all $t \in \tau$ we define $\epsilon_t^{i} = \chi_{i,t} - \left[ W_{(T-\tilde{t})} B \right]_{i,t}.$
}
\end{enumerate}
}
\normalfont
The proof for the conjugacy for Conditional 1 and Conditional 2 are provided in \cite{Sugita2002} [Section 2.2, Equations 10 and 11] as a direct consequence of Theorem 1 and Theorem 2 and the transformation developed and conjugate prior choices. The derivation of Conditional 3 also follows from \cite{Sugita2002} [Section 2.2, Equation 14]. The proof for Conditional 4 is presented in \cite{godsill00} [Section 2 Equation 4]. We will later demonstrate in Section \ref{modelCVAR3} how this conjugacy can be beneficially utilized as a proposal distribution in a ABC general non-symmetric $\alpha$-stable Bayesian CVAR model. In addition we will provide novel algorithms to sample from the resulting posterior distributions also in Section \ref{modelCVAR3}.

\subsection{General $\alpha$-stable Approximate Bayesian Computation CVAR Model} \label{GeneralStableModel}
Under the noise model presented in Equation (\ref{error}) we have an intractable matrix-variate likelihood model since the asymmetric $\alpha$-stable inter-day model does not admit a density. Hence, our noise model for the i-th series at time t becomes,
\begin{equation}
\begin{split}
\epsilon^{(i)}_t &\sim \mathcal{N}\left(0,\sigma^{(i)}\right) \mathbb{I}\left(t \notin \bm{\tau}\right) + \mathcal{S}_{a^{(i)}}\left(b^{(i)}, \gamma^{(i)}, \delta^{(i)}\right) \mathbb{I}\left(t \in \bm{\tau}\right). 
\label{errorGeneralStable}
\end{split}
\end{equation}
In this section we develop an ABC model and associated Markov chain Monte Carlo (MCMC-ABC) sampler to perform estimation in this general composite CVAR noise model setting. MCMC-ABC samplers are actively studied in the statistical literature since \cite{tavare+bgd97}, see a review chapter in \cite{sisson-likelihood}.

ABC inference adopts the approach of augmenting the target posterior distribution from the intractable ``True'' model, denoted $p(\Sigma, B, \bm{\beta}|Y) \propto p(Y|\Sigma, B, \bm{\beta})p(\Sigma, B, \bm{\beta})$, into an augmented target posterior distribution. The ABC posterior model approximation, denoted $p_{ABC}\left(\Sigma, B, \bm{\beta}|Y\right)$, is therefore defined by,
\begin{equation}
p_{ABC}\left(\Sigma, B, \bm{\beta},Y_S|Y\right) 
= p(Y|Y_S,\Sigma, B, \bm{\beta}) p(Y_S|\Sigma, B, \bm{\beta})  p(\Sigma, B, \bm{\beta})   
\end{equation}
where the auxiliary parameters ``synthetic observation'' matrix $Y_S$ are a (simulated) dataset from $p(Y_S|\Sigma, B, \bm{\beta})$, on the same space as $Y$. The function $p(Y|Y_S,\Sigma, B, \bm{\beta})$ is chosen to weight the posterior $p(\Sigma, B, \bm{\beta}|Y)$ with high values in regions where $Y_S$ and $Y$ are similar. There are many choices for this function discussed and studied in \cite{peters2010chain} but generally it is assumed to be constant with respect to parameters $\Sigma, B, \bm{\beta}$ at the point $Y_S = Y$, so that $p(Y|Y,\Sigma, B, \bm{\beta}) = c$, for some constant $c>0$, with the result that the target posterior is recovered exactly at  $Y_S = Y$. That is $p_{ABC}\left(\Sigma, B, \bm{\beta}, Y|Y\right) =p(\Sigma, B, \bm{\beta}|Y)$

Given the augmented ABC posterior distribution  $p_{ABC}\left(\Sigma, B, \bm{\beta}, Y_S|Y\right)$ generally inference involves the marginal posterior,
\begin{equation}
p_{ABC}\left(\Sigma, B, \bm{\beta}|Y\right) 
\propto p(\Sigma, B, \bm{\beta}) \int p(Y|Y_S,\Sigma, B, \bm{\beta}) p(Y_S|\Sigma, B, \bm{\beta}) dY_S     
\end{equation}
obtained by integrating out the auxiliary dataset. The ABC distribution $p_{ABC}\left(\Sigma, B, \bm{\beta}|Y\right) $ then acts as an approximation to $p\left(\Sigma, B, \bm{\beta}|Y\right)$ and is obtained in practice by discarding realizations of the auxiliary dataset from the output of any sampler targeting the joint posterior $p_{ABC}\left(\Sigma, B, \bm{\beta}, Y_S|Y\right)$.

Generally, the weighting function $p(Y|Y_S,\Sigma, B, \bm{\beta})$ is simplified in two important ways, the first involves replacing the observation and synthetic data vector / matrix with summary statistics and the second involves making a kernel approximation to the weighting function. Therefore we obtain a kernel representation of the form
$$p_{\epsilon}(Y|X,\Sigma, B, \bm{\beta}) = \frac{1}{\epsilon}K \left( \frac{|\bm{S}(X)-\bm{S}(Y)|}{\epsilon}\right),$$
see \cite{peters2010chain}, \cite{ratmann+ahwr09} and \cite{beaumont2009adaptivity}. In this simplification the data matrix $Y$ is replaced with summary statistics (ideally sufficient statistics) vector or matrix denoted $\bm{S}(Y)$ of significantly lower dimension than $Y$. When sufficient statistics are not available, then summary statistics are utilized at the cost of bias, see recent discussion in \cite{fearnhead2010semi}. 

We consider a hard decision kernel weighting function (uniform kernel) with Euclidean L2-norm distance measure between summary statistics on vectorized observation matrices $Vec(Y)$ and $Vec(Y_S)$ given by 
\begin{equation*}
\label{weightABC}
p_{\epsilon}(Y|Y_S,\Sigma, B, \bm{\beta}) = 
\left\{\begin{array}{cc}
1 & \text{if } ||\bm{S}(Vec(Y)) - \bm{S}(Vec(Y_S))|| \leq \epsilon \\
 0 &  \text{otherwise}\\
\end{array}\right.
\end{equation*}
\textbf{Remark 6:}  \textit{For sufficient statistics and as $\epsilon \rightarrow 0$ it has been proven that an MCMC-ABC sampler with this kernel, will obtain correlated samples from the stationary regime given by the target posterior distribution $p(\Sigma, B, \bm{\beta}|Y)$, see a review in \cite{sisson-likelihood}.}\\
\noindent \textbf{Remark 7:}  \textit{The model we propose is highly non-standard in the ABC literature since it involves a combination of likelihood components some of which are tractable and others which are intractable. This opens the possibility of many alternative sampling approaches, for example we could compute the likelihood for the tractable portions of time and then approximate the likelihood for the portions of time in which the noise model produces an intractable likelihood.}

The particular algorithm we consider in Section \ref{modelCVAR3} will demonstrate how to combine both the SMiN and ABC Bayesian CVAR models developed. In particular providing a general adaptive MCMC based sampling algorithm for matrix variate $\alpha$-stable CVAR posterior distributions in the approximate Bayesian computation setting. This involves use of the conjugate models derived under the SMiN assumption as proposal distributions in the ABC sampler, reducing the required dimension of the adaptive proposal kernel in our MCMC sampler.

Hence we have developed two novel Bayesian modeling frameworks for incorporation of the $\alpha$-stable model in the CVAR model framework. We can now consider inference and sampling under these models.
\section{Sampling and Estimation}
\label{modelCVAR3}

Here we focus on obtaining samples from the matrix variate posterior distributions derived in Section \ref{GeneralStableModel}. We will achieve this via design of a novel sampling methodology we develop based on adaptive MCMC in a ABC setting. It is a hybrid algorithm since the proposal distribution for several of the posterior matrix variables ($\Sigma$,$\tilde{B}$) in the ABC sampling framework are sampled via the conjugate model derived for the symmetric $\alpha$-stable case in Theorem 3, which in this case acts as a proposal for the non-symmetric model in the ABC framework. The remaining matrix posterior parameters ($\bm{\beta}$,$\bm{\lambda}$) are sampled via an adaptive Metropolis and adaptive Rejection Sampling framework. The proposal are combined into the ABC methodology as presented in Algorithm 1.

\subsection{Hybrid Adaptive Markov Chain Monte Carlo ABC.}
Here we present the sampling methodology for posterior $p_{ABC}(\Sigma,\bm{B},\bm{\beta}|\bm{Y})$ conditional on estimation of the $\alpha$-stable inter-day parameters on the batch of data $\bm{Y}$ under consideration. Note below we present a version of the HAdMCMC-ABC algorithm in which all matrix parameters are updated at each iteration of the Markov chain, however, block Metropolis-within-Gibbs frameworks are trivial to also consider. The resulting proposal distribution for the MCMC sampler comprises a hybrid proposal comprised of a conjugate posterior proposal under the symmetric $\alpha$-stable setting and an adaptive Metropolis proposal. Proposing to update the matrix variate Markov chain parameters from iteration $j-1$ to iteration $j$ involves sampling proposal $\left\{\Sigma,\bm{B},\bm{\beta},\lambda \right\}$ given Markov chain state $\left\{\Sigma,\bm{B},\bm{\beta},\lambda \right\}[j-1]$ according to the proposal,
\begin{equation}
\begin{split}
q &\left( \left\{  \Sigma, \bm{B}, \bm{\beta}, \lambda \right\} ; \left\{ \Sigma, \bm{B}, \bm{\beta}, \lambda \right\} \right) \\
&= p(\Sigma|\bm{\beta},\bm{\lambda},\bm{\alpha},\bm{\gamma},\bm{\delta},\widetilde{Y}) p(\tilde{B}|\bm{\beta},\bm{\lambda},\bm{\alpha},\bm{\gamma},\bm{\delta},\Sigma,Z_*,Q_*)
p(\lambda_i|\bm{\alpha},\bm{\gamma},\bm{\delta},\chi,\tilde{\bm{B}},Q_*,\bm{\beta})
q(\bm{\beta},\bm{\beta}[j-1])
\end{split}
\end{equation}
where the first three proposal distributions for the Markov chain are given by Theorem 3 under a symmetric $\alpha$-stable intra-day assumption allowing them to be sampled exactly and $q(\bm{\beta},\bm{\beta}[j-1])$ is given by the adaptive Metropolis proposal developed in \cite{peters2010model} [Algorithm 2]. 

{\footnotesize{
\begin{algorithm}
\dontprintsemicolon
\KwIn{ Initialized Markov chain matrix variate states $\bm{\theta}^{(0)}=\left(\Sigma^{(0)}, \tilde{B}^{(0)}, \bm{\beta}^{(0)}, \bm{\lambda}^{(0)}\right)$.}
\KwOut{Markov chain samples
$\{\bm{\theta}^{(j)}\}_{j=1:J} = \{\Sigma^{(j)},B^{(j)},\bm{\beta}^{(j)}\}_{j=1:J} \sim
p_{ABC}\left(\Sigma,B,\bm{\beta}|Y\right)$.}
\Begin{
\begin{enumerate}
\item[1a.] Set ABC tolerance level $\epsilon$ (note annealing of the tolerance can be utilized). \;
\item[1b.] Evaluate summary statistic vector for observed price series vectors $\bm{S}(Vec(\bm{Y}))$. \;
\end{enumerate}
\Repeat{j = J}{
\begin{enumerate}
\item[2.] Sample conjugate proposals for matrix parameters ($\Sigma$,$\tilde{B}$):
\begin{enumerate}
\item[2a.] Sample proposed matrix state $\Sigma^{*}$ via inversion from conjugate posterior $p(\Sigma|\bm{\beta}^{(j-1)},\bm{\lambda}^{(j-1)},\bm{\alpha}^{(j-1)},\bm{\gamma},\bm{\delta},\widetilde{Y})$,  [Theorem 1: Conditional 1]. 
\item[2b.] Evaluate transformation matrix $Q_*^{*}$ based on proposed state $\Sigma^{*}$ and obtain transformed observation matrix $Z_*$, [Lemma 3].
\item[2c.] Sample proposed matrix state $\bm{\tilde{B}}^{*}$ via inversion from $p(\tilde{B}|\bm{\beta}^{(j-1)},\bm{\lambda}^{(j-1)},\bm{\alpha}^{(j-1)},\bm{\gamma},\bm{\delta},\Sigma^{*},Z_*,Q_*^{*})$, [Theorem 1: Conditional 2]. 
\end{enumerate}
\item[3.] Sample adaptive proposals for matrix parameters ($\bm{\beta}$,$\bm{\lambda}$):
\begin{enumerate}
\item[3a.] Sample proposed vector $\bm{\lambda}^{*}$ with each component sampled from $p(\lambda_i|\bm{\alpha},\bm{\gamma},\bm{\delta},\chi,\tilde{\bm{B}},Q_*,\bm{\beta})$, in [Theorem 1: Conditional 4] via single component adaptive rejection sampling proposed in Godsill (2000) [Section 3.1.1., p.2]
\item[3b.] Sample proposed unconstrained elements of matrix $\bm{\beta}$ from adaptive metropolis proposal in Peters et al. (2010) [Algorithm 2, p.12].
 
\end{enumerate}

\item[4.] Generate synthetic data set $\bm{Y}_S$ given proposal $\left(\Sigma, B, \beta, \bm{\lambda}\right)$ and fitted intra-day model $S_{\bm{\alpha}}(\bm{\beta},\bm{\gamma},\bm{\delta})$:
\begin{enumerate}
\item[4a.] Evaluate summary statistic vector for synthetically generated price series vectors $\bm{S}(Vec(\bm{Y}_S))$. 
\item[4b.] Calculate weighting function in Equation \ref{weightABC}. 
\end{enumerate}

\end{enumerate}

\begin{enumerate}
\item[5.]{Calculate ABC - Metropolis Hastings Acceptance Probability according to the general specification in Sisson and Fan (2010) [Equation 1.3.2] for joint proposal $\bm{\theta} = \left(\Sigma, B, \beta, \bm{\lambda}\right)$:
\begin{equation}
A\left(\bm{\theta}^{(j-1)},\bm{\theta}^{*}\right) = \frac{p_{ABC}\left(\bm{\theta}^{*}|Y\right)q\left(\bm{\theta}^{*} \rightarrow \bm{\theta}^{(j-1)}\right) }
{p_{ABC}\left(\bm{\theta}^{(j-1)}|Y\right)q\left(\bm{\theta}^{(j-1)} \rightarrow \bm{\theta}^{*}\right)}
\end{equation}
Accept $\bm{\theta}^{(j)}=\bm{\theta}^*$ via rejection using A, otherwise $\bm{\theta}^{(j)}=\bm{\theta}^{(j-1)}$. Set j = j+1.}\;
\end{enumerate}
}
}
\caption{Hybrid Adaptive Markov Chain Monte Carlo Approximate Bayesian Computation (HAdMCMC-ABC).\label{HAdMCMC-ABC}}
\end{algorithm}
}}

\section{Results and Analysis}
In this section we perform three studies. The first part involves numerical analysis of the algorithms developed to sample from the matrix variate posterior distribution on data sets generated with known parameters. This is performed in two settings, the first under a mixture noise model (Equation \ref{errorMixture}) with very heavy tailed symmetric $\alpha$-stable inter-day noise ($\alpha = 1.3$). In this case according to Theorem 3, we know the exact posterior full conditional distributions. Therefore, sampling results from this model are compared for the resulting exact MCMC sampler, denoted ''Mixture Exact'' versus a ABC approximation sampler generated under the ABC approximate model sampled via Algorithm 1, denoted ''Mixture ABC''. In addition, we ignore the stable innovations and run the adaptive MCMC sampler and posterior models of \cite{peters2010model} and \cite{sugita2009monte}, denoted (''Gaussian''), to further assess the bias in parameter estimates if intra-day level shifts are not modeled explicitly. We are particularly interested in the estimated cointegration basis vector $\beta$ which directly affects portfolio weights in pairs trading settings.

The second study considers asymmetric heavy tailed $\alpha$-stable ($\alpha = 1.3$, $\beta = 0.5$) inter-day noise. In this case we can only compare the ABC model and MCMC results from Algorithm 1 to the case in which intra-day level shifts are ignored  in the ''Gaussian'' case and sampling occurs as in \cite{peters2010model}. In the third study we consider a real data set analysis via our general MCMC-ABC sampler in Algorithm 1, for a pair of assets, observed in practice to have a cointegration relationship with rank $r=1$, with $\alpha$-stable fits from Table \ref{tab:AlphaStableMLEfits} for $AUD - CD$.

In all studies we consider pairs data, with a cointegration rank of $r=1$. We ran samplers with 10,000 burn in samples and 20,000 actual samples. In studies one and two we perform analysis on 20 independently generated pairs of price data sets, with each price series of length 500 samples and every 50-th sample modeled with an $\alpha$-stable innovation. In the real data analysis we take the series described in Section 4.1.

\subsection{Synthetic Data Analysis - Symmetric Case}
The model used for this synthetic study considers parameter settings $\beta = [1, 0.5]$, $\alpha = [0.1, -0.3]$, $\Sigma = \mathbb{I}_2$ $\mu = [0, 0]$ and ($\alpha = 1.3$, $\beta = 0$, $\gamma = 1$, $\delta = 0$). The prior settings for the Bayesian model are those specified in \cite{peters2010model}. The ABC tolerance level used was $\epsilon = 0.1$. In Table 2 we present the results comparing the performance of the estimation of the parameters for the resulting Bayesian posterior model in Theorem 3.
\begin{table}
{\centering {\small{
\begin{tabular}{|c|c|c|c|c|}
    \hline
      & \textbf{Gaussian model} & \multicolumn{3}{|c|}{\textbf{Mixture Gaussian and $\alpha$-stable intra-day model}}  \\ \hline
    \textbf{Parameter Estimates} & \textbf{Gaussian} & \textbf{Mixture ABC} & \textbf{Mixture Exact} & \textbf{Truth} \\ \hline
Ave. MMSE $\beta_{1,2}$   & -0.02 (0.21) & 0.39 (0.27) 	& 0.42 (0.25) & 0.5 \\ \hline
Ave. Stdev. $\beta_{1,2}$ & 0.28 (0.08)  & 0.31 (0.12)	& 0.35 (0.09) & -\\ \hline
Ave. MMSE $\text{tr}\left(\Sigma\right)$
                          & 3.17 (2.03)  & 2.61 (2.12)	& 2.23 (1.91) &2\\ \hline
Ave. Stdev. $\text{tr}\left(\Sigma\right)$
                          & 0.16 (0.12)  & 0.21 (0.16)	& 0.19 (0.21) &- \\ \hline
Ave. MMSE $\mu_1$         & -0.03 (0.08) & -0.01 (0.03)	& 0.05 (0.01) &- \\ \hline
Ave. Stdev. $\mu_1$       & 0.06 (0.03)  & 0.08 (0.02)	& 0.07 (0.02) &-\\ \hline
Ave. MMSE $\mu_2$         & 4.0E-3 (0.01)& 7E-3 (0.03)	& 6E-3 (0.01) &0.1\\ \hline
Ave. Stdev. $\mu_2$       & 0.05 (0.01)  & 0.07 (0.02)	& 0.09 (0.03) & - \\ \hline
Ave. MMSE $\alpha_{1,1}$  & -0.06 (0.02) & 0.05 (0.02)	& 0.08 (0.04) &0.1 \\ \hline
Ave. Stdev. $\alpha_{1,1}$& 0.02 (2E-3)  & 0.03 (4E-3)	& 0.05 (3E-3) & -\\ \hline
Ave. MMSE $\alpha_{1,2}$  & 3E-3 (0.02)  &-0.19 (0.01)& -0.21 (0.02) &-0.3 \\ \hline
Ave. Stdev. $\alpha_{1,2}$& 0.02 (0.01)  & 0.02 (0.01)	& 0.04 (0.02) & -\\ 
\hline\hline
Ave. Mean acceptance probability & 0.37 & 0.21 & 1& -\\ \hline
\end{tabular}
\caption{{\textbf{Sampler Analysis: }} Ave. MMSE or Stdev is averaged posterior mean or variances obtained from estimation of the posterior parameters from 20 independently generated data sets. In $(\cdot)$ are the standard error in estimates. In all simulations the initial Markov chain is started far away from the true parameter values.} }}
\centering} 
\end{table}
The results demonstrate that the effect of ignoring the inter-day level shifts when fitting the Bayesian model has a significant effect on the estimation of the cointegration vector $\bm{\beta}$. In addition, it is clear that in this symmetric case, the estimates obtained via the exact MCMC sampler and the ABC approximation are similar. However, as expected, the computational cost for the ABC approach is significantly higher than the non-ABC approach. We also see that estimation of the other parameters are also accurate. We summarize the results for the cointegration vector $\bm{\beta}$ of the estimated MMSE in Figure 4 under the Gaussian case ignoring the intra-day level shifts and the mixture model proposed in this paper.

\subsection{Synthetic Data Analysis - Asymmetric Case}
The model used for this synthetic study considers identical parameter settings and prior settings for the CVAR model as the previous study, with the asymmetric inter-day noise model with $\alpha$-stable parameters ($\alpha = 1.3$, $\beta = 0.5$, $\gamma = 1$, $\delta = 0$). The ABC tolerance level used was $\epsilon = 0.1$. In the asymmetric case we must work with the ABC Bayesian model. In Table 3 we present the results comparing the performance of the estimation of the parameters for the resulting ABC Bayesian posterior versus the basic Gaussian conjugate Bayesian model.
\begin{table}
{\centering {\small{
\begin{tabular}{|c|c|c|c|}
    \hline
      & \textbf{Gaussian model} & \multicolumn{2}{|c|}{\textbf{Mixture Gaussian and $\alpha$-stable intra-day model}}  \\ \hline
    \textbf{Parameter Estimates} & \textbf{Gaussian} & \textbf{Mixture ABC} & \textbf{Truth} \\ \hline
Ave. MMSE $\beta_{1,2}$   & -0.01 (0.21) & 0.36 (0.32) 	& 0.5 \\ \hline
Ave. Stdev. $\beta_{1,2}$ & 0.28 (0.08)  & 0.41 (0.16)  & -\\ \hline
Ave. MMSE $\text{tr}\left(\Sigma\right)$
                          & 2.92 (1.32)  & 3.0 (1.49)	&2\\ \hline
Ave. Stdev. $\text{tr}\left(\Sigma\right)$
                          & 0.14 (0.07)  & 0.21 (0.12)	&- \\ \hline
Ave. MMSE $\mu_1$         & -0.02 (0.07) & -0.01 (0.09)	&0.1 \\ \hline
Ave. Stdev. $\mu_1$       & 0.06 (0.02)  & 0.10 (0.03)	&-\\ \hline
Ave. MMSE $\mu_2$         & -3.0E-3 (0.01)& 4E-3 (0.03)	&0.1\\ \hline
Ave. Stdev. $\mu_2$       & 0.05 (0.01)  & 0.09 (0.03)	& - \\ \hline
Ave. MMSE $\alpha_{1,1}$  & -0.06 (0.01) & 0.06 (0.03)	&0.1 \\ \hline
Ave. Stdev. $\alpha_{1,1}$& 0.01 (2E-3)  & 0.03 (8E-3)	& -\\ \hline
Ave. MMSE $\alpha_{1,2}$  & 2E-3 (0.02)  & 1E-3 (8E-3)  &-0.3 \\ \hline
Ave. Stdev. $\alpha_{1,2}$& 0.02 (0.01)  & 0.03 (0.01)	& -\\ 
\hline\hline
Ave. Mean acceptance probability & 0.42 & 0.28 & -\\ \hline
\end{tabular}
\caption{{\textbf{Sampler Analysis: }} Ave. MMSE or Stdev is averaged posterior mean or variances obtained from estimation of the posterior parameters from 20 independently generated data sets. In $(\cdot)$ are the standard error in estimates. In all simulations the initial Markov
chain is started very far away from the true parameter values.} }}
\centering} 
\label{SyntheticAsymmetricResults}
\end{table}
Estimation results in Table 3 demonstrate significantly more accurate results for the estimation of the cointegration vectors when inter-day noise modeling is incorporated. Again, we summarize the results for the cointegration vector $\bm{\beta}$ of the estimated MMSE in Figure 4 under the Gaussian case ignoring the inter-day level shifts and the mixture model proposed in this paper.
\begin{figure}[!h]
\label{fig:DIFF}
\centering
\includegraphics[height=0.2\textheight,width=0.45\textwidth]{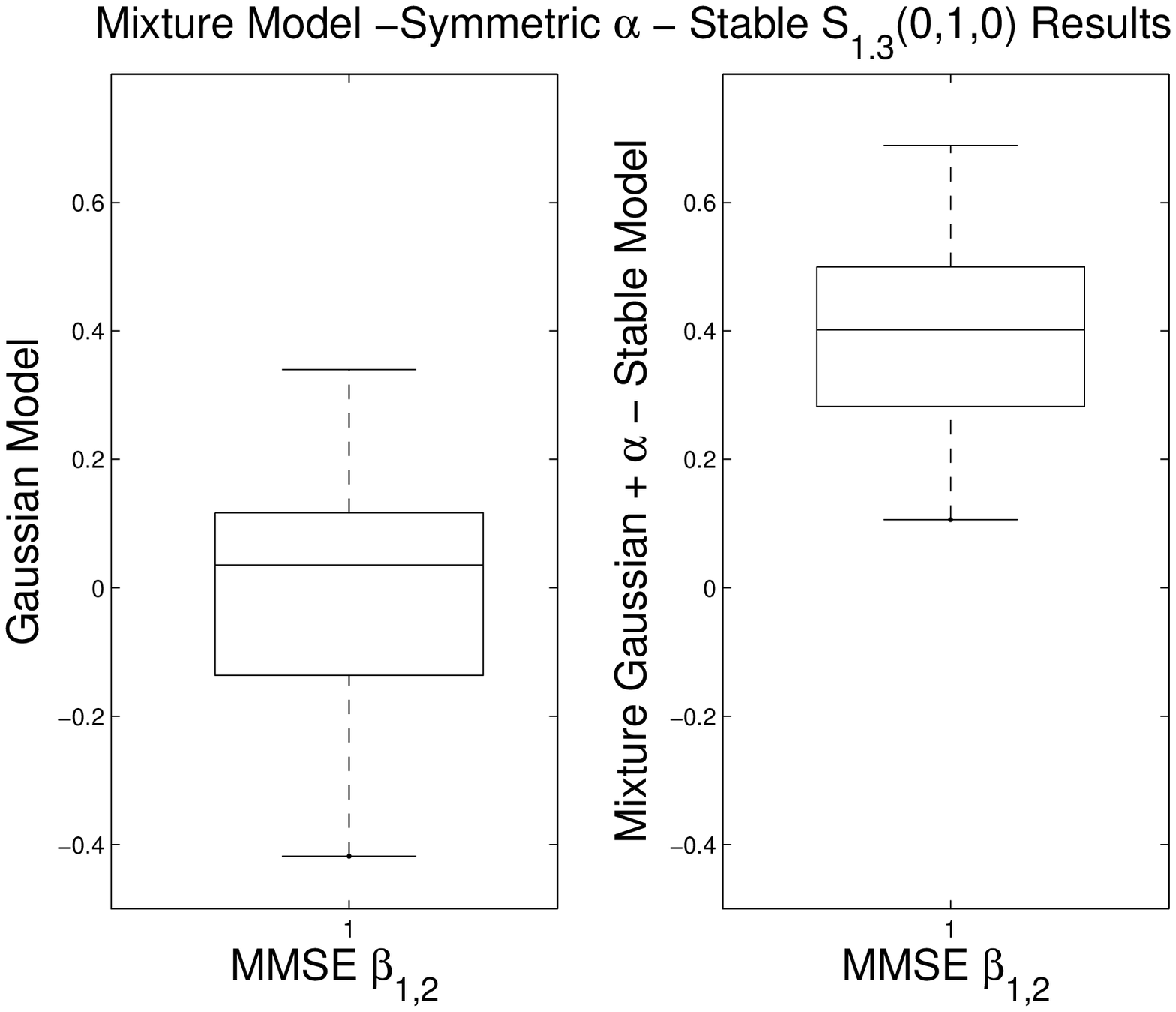}
\includegraphics[height=0.2\textheight,width=0.45\textwidth]{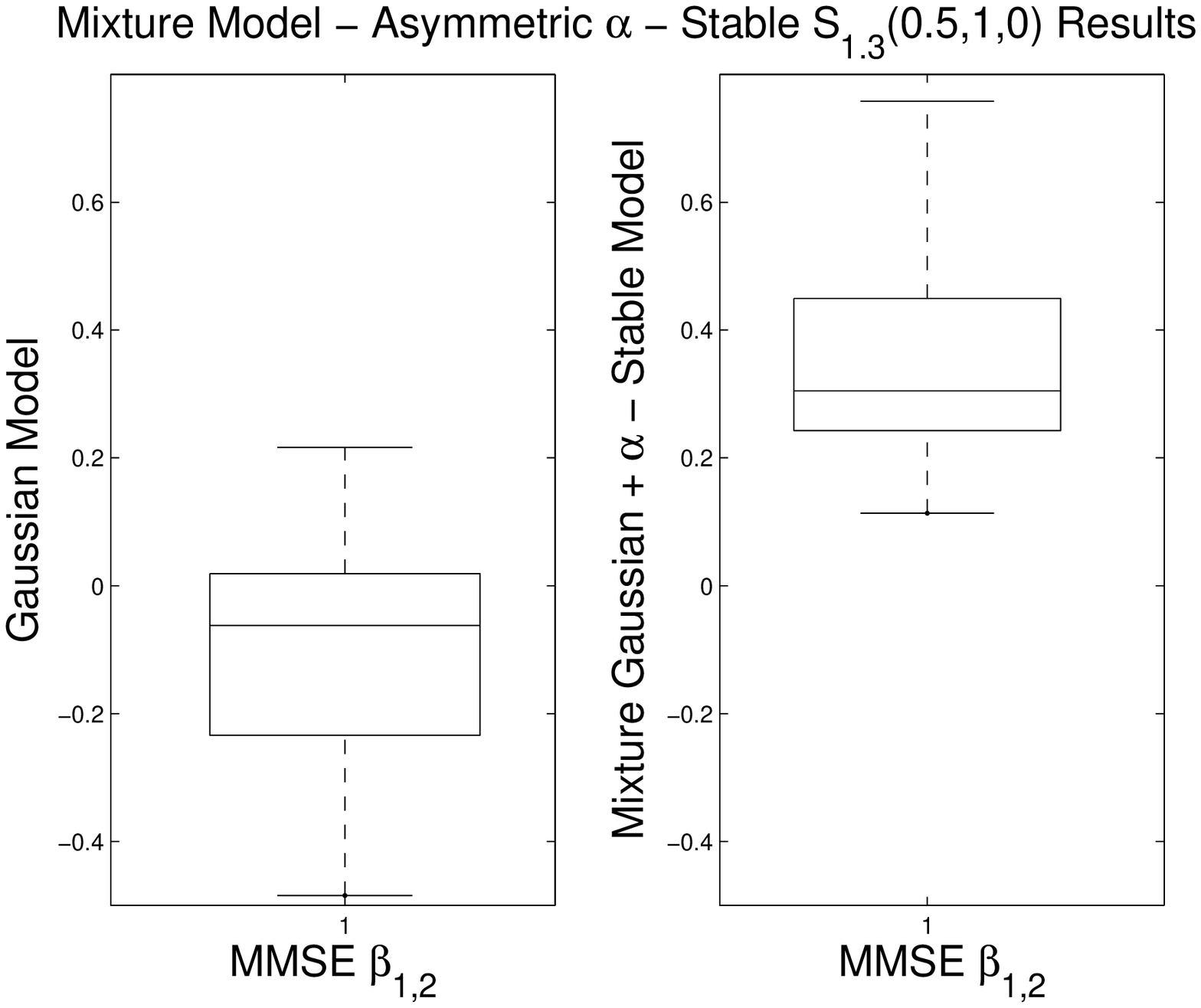}
\caption{Estimated cointegration vector $\bm{\beta}$.}
\end{figure}

\subsection{Real Data Analysis}
In this section we particularly focus on the accuracy of estimation of the cointegration vectors $\bm{\beta}$. These are important to the design of algorithmic trading strategies since they are the basis for projection of the raw price series to obtain a stationary deviation series to consider trading analysis. In addition we provide estimation results for the reversion rate of the stochastic trends to stationarity as denoted by the matrix $\bm{\alpha}$. We analyze the performance of the basic ''Gaussian'' posterior model of \cite{peters2010model} and \cite{sugita2009monte} in the presence of inter-day price series level shifts versus the estimation of the ''Mixture ABC'' model via Algorithm 1. 

The price series for AUD / CD with base currency in AUD sampled at 10min intervals during the joint open market hours. Analysis is performed for the first contract in Table 1, starting from the 05/09/99, containing 60 days worth of market data, producing a time series of prices of length 29,621 samples. The raw price series are presented in Figure 5 with circles representing the joint open of each market (inter-day boundaries).
\begin{figure}[!h]
\label{fig:PriceSeriesReal}
\centering
\includegraphics[height=0.3\textheight,width=0.9\textwidth]{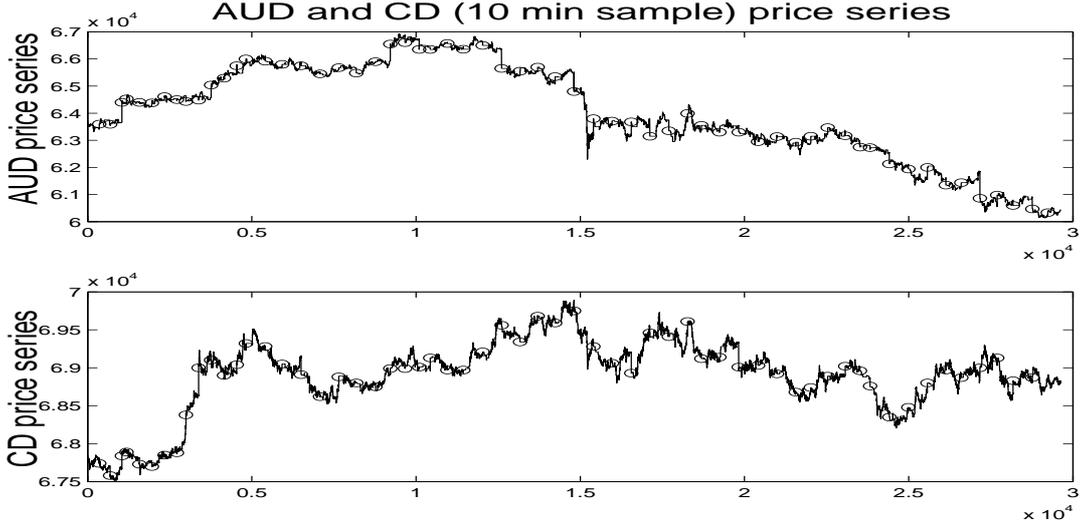}
\caption{Price series for AUD and CD. Circles indicate inter-day market time boundaries.}
\end{figure}
The data was transformed by translation of each series by the median and scaled by the standard deviation. The analysis performed considers 30 batches of 2 days of data, giving on average 489 data samples per batch, and the posterior parameter estimates are averaged over samplers analysis of each data set and presented in Table 4.
\begin{table}
{\centering {\small{
\begin{tabular}{|c|c|c|}
    \hline
      & \textbf{Gaussian model} & \multicolumn{1}{|c|}{\textbf{Mixture Gaussian and $\alpha$-stable intra-day model}}  \\ \hline
    \textbf{Parameter Estimates} & \textbf{Gaussian} & \textbf{Mixture ABC} \\ \hline
Ave. MMSE $\beta_{1,2}$   & -0.31 (0.25) 			& 0.18 (0.21) 	\\ \hline
Ave. Var. $\beta_{1,2}$ 	& 0.20  (0.04) 			& 0.83 (0.08)  	\\ \hline
Ave. MMSE $\alpha_{1,1}$  & -0.02 (1.36E-3) 	& -0.01 (3.8E-3)		\\ \hline
Ave. Var. $\alpha_{1,1}$	& 3.90E-5 (4.09E-6) & 5.3E-5 (2.5E-5)		\\ \hline
Ave. MMSE $\alpha_{1,2}$  & 1.24E-3(1.20E-3) 	& -6.3E-4 (1.7E-3)  	\\ \hline
Ave. Var. $\alpha_{1,2}$	& 2.18E-5(3.07E-6) 	& 1.7E-5 (1.0E-3)		\\ 
\hline
\end{tabular}
\caption{{\textbf{Sampler Analysis: }} In $(\cdot)$ are the standard error estimates obtained from 20 batches of MCMC samples each of length 1,000, average over each of the sets of 2 days of data.} }}
\centering} 
\label{RealDataAUDCD}
\end{table}
The results demonstrate that failing to account for the inter-day level shifts observed can significantly affect the estimation of the cointegration vectors and reversion rates as demonstrated in the comparison in Table \ref{RealDataAUDCD} and in Figure \ref{fig:BETAREAL}.
\begin{figure}[!h]
\label{fig:BETAREAL}
\centering
\includegraphics[height=0.3\textheight,width=0.9\textwidth]{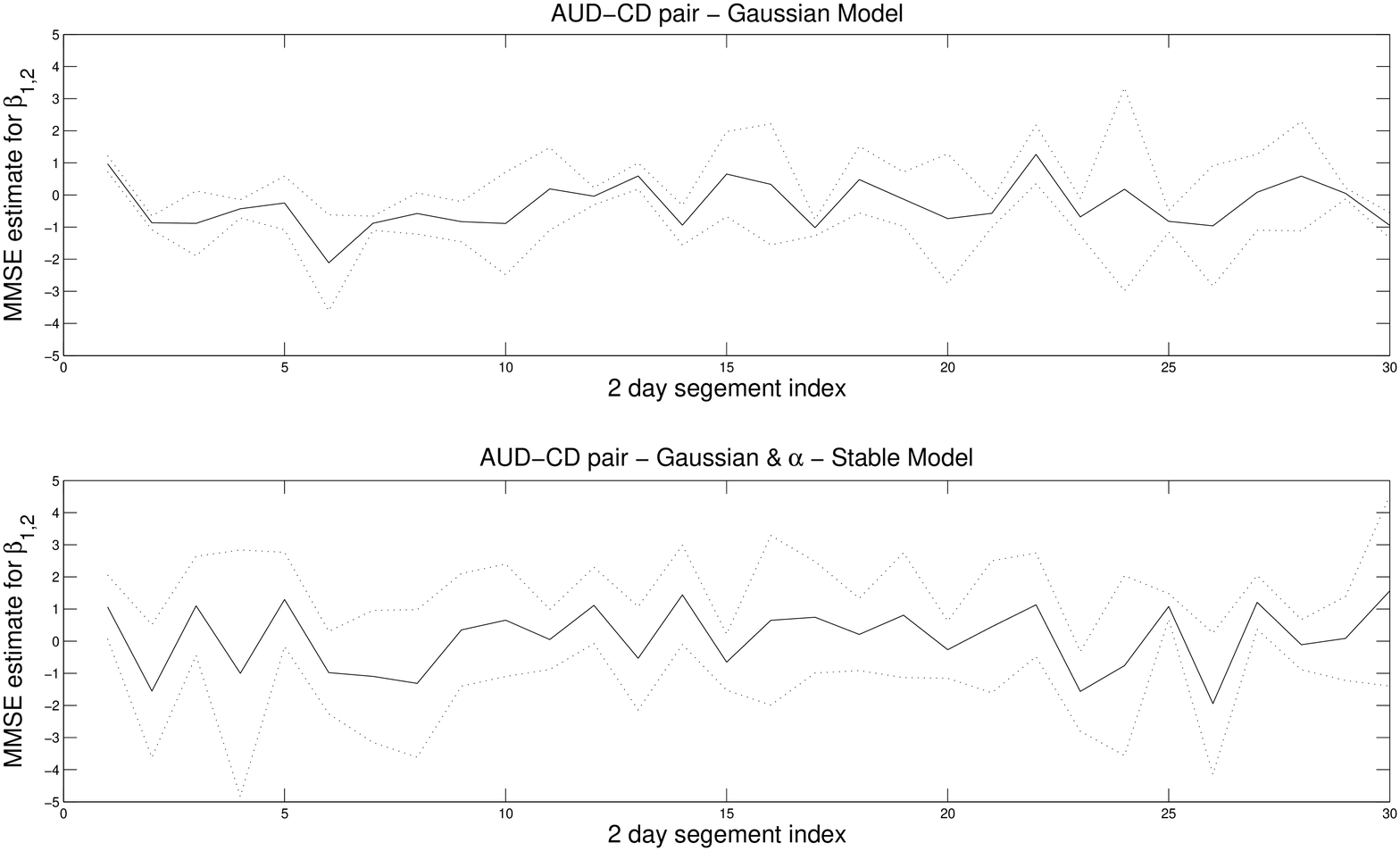}
\caption{Estimated cointegration vector $\bm{\beta}$ for AUD-CD pair for 2 day segments at 10min samples. TOP: Gaussian model; Bottom: Mixture ABC model; Solid line is estimated MMSE and dashed line is posterior 95\% C.I..}
\end{figure}

\section{Conclusions}
We studied the impact of price series level shifts on statistical estimation of matrix variate parameters in CVAR models utilized in algorithmic trading. In particular we first demonstrated the significant impact on estimation accuracy under both frequentist and Bayesian estimation frameworks when failing to appropriately model observed level shifts in price series. 

Next we developed a composite noise model comprised of Gaussian and $\alpha$-stable innovation noise for the CVAR model in the presence of price series level shifts. The example we illustrated this point on was the situation that occurs at deterministic times each trading day, at the inter-day market boundaries. However, we point out that our methodology is general and extends also to settings in which the level shift times are unknown \textit{a priori}. This would modify the problem to additional estimation of the $\tau$ times, then conditional on these estimates, our methodology can be applied.

Working under this composite noise model of Gaussian and $\alpha$-stable CVAR innovations, we developed a novel conjugate Bayesian model under transformation, allowing for exact MCMC sampling frameworks to be developed in the symmetric heavy tailed $\alpha$-stable scenario. In the asymmetric skewed noise setting, a non-standard approximate Bayesian computation model was developed and an advanced adaptive MCMC algorithm was utilized to sample this ABC posterior. This incorporated the conjugate model developed in the symmetric case as an MCMC-ABC proposal for the asymmetric setting. 

We were able to demonstrate and verify on synthetic data sets under both symmetric $\alpha$-stable and asymetric $\alpha$-stable models, that the sampling methodology we developed for estimation of the MMSE for the matrix variate posterior parameters is accurate. We then compared the performance of our model and sampler to the standard Gaussian Bayesian CVAR model on real financial pairs, demonstrating a marked difference in the estimated CVAR model parameters. Hence, justifying the applicability of such a model in applied financial models for trading.

Our framework was motivated from the perspective that our approach is justified by the assumption that the underlying model for the price series pair is appropriately modeled by the basic CVAR model presented in Section \ref{modelCVAR}. This differs significantly to the underlying assumption of \cite{chen2010subsampling}. If this assumption is not suitable, alternative approaches could be considered, such as the use of a Markov switching regime model, see for example \cite{krolzig1997statistical}. Under such a model the CVAR parameters may vary depending on a latent regime state variable, see \cite{sugita2008bayesian} for details. These models are suitable in settings in which one believes there is fundamentally a finite set of distinct models suitable for describing the statistical properties of the vector price series. In such settings, typically the parameters of each model and the transition times for model switching are unknown and must be estimated. The model framework we present is distinctly different to this setting, not only do we know the deterministic times at which level shifts in the price series occur at the open and close of markets, but we also assume after accounting for these, the fundamental CVAR model parameterizes appropriately the underlying assets price series. 

Other possible extensions that can be made to your model are considerations of mixtures of Student-t and Gaussian innovation errors for intra-day innovation noise. This would allow one to capture possible skew or heavy tailedness present withing the trading day in certain markets, whilst still maintaining our conjugacy properties. 

\section{Acknowledgements}
We thank Prof. Arnaud Doucet and Prof. Richard Gerlach for comments and suggestions during this research. In addition, we thank the Institute of Statistical Mathematics, Tokyo, Japan, for hosting GWP whilst performing aspects of this research and Boronia for funding.

\bibliographystyle{plainnat}
\bibliography{JCF}

%
%
%
%
%
%
%
%
%
%

\end{document}